%% file: main.tex
\documentclass[twocolumn]{aastex62}

\usepackage{longtable}
\usepackage{amssymb,amsmath}
\usepackage{array,booktabs}
\usepackage{graphicx}
\usepackage{multirow}
\usepackage{hyperref}
\usepackage{color}
\usepackage{soul}
\usepackage{rotating}
\usepackage{mhchem}

\definecolor{darkgreen}{rgb}{0.0,,0.0}

\shorttitle{}
\shortauthors{}

\begin{document}

\title{Obliquity Variations of Habitable Zone Planets Kepler-62f and Kepler-186f}

\correspondingauthor{Yutong Shan}\email{yshan@cfa.harvard.edu}

\author{Yutong Shan}\affiliation{Harvard-Smithsonian Center for Astrophysics, 60 Garden Street, Cambridge, MA 02138, USA }

\author{Gongjie Li}\affiliation{Harvard-Smithsonian Center for Astrophysics, 60 Garden Street, Cambridge, MA 02138, USA }\affiliation{Center for Relativistic Astrophysics, School of Physics, Georgia Institute of Technology, Atlanta, GA 30332, USA} 

\begin{abstract}
Obliquity variability could play an important role in the climate and habitability of a planet. Orbital modulations caused by planetary companions and the planet's spin axis precession due to the torque from the host star may lead to resonant interactions and cause large-amplitude obliquity variability. Here we consider the spin axis dynamics of Kepler-62f and Kepler-186f, both of which reside in the habitable zone around their host stars. Using {\emph{N}}-body simulations and secular numerical integrations, we describe their obliquity evolution for particular realizations of the planetary systems. We then use a generalized analytic framework to characterize regions in parameter space where the obliquity is variable with large amplitude. We find that the locations of variability are fine-tuned over the planetary properties and system architecture in the lower-obliquity regimes ($\lesssim 40^\circ$). 
As an example, assuming a rotation period of 24 hr, the obliquities of both Kepler-62f and Kepler-186f are stable below $\sim 40^\circ$, whereas the high-obliquity regions ($60^\circ - 90^\circ$) allow moderate variabilities. However, for some other rotation periods of Kepler-62f or Kepler-186f, the lower-obliquity regions could become more variable owing to resonant interactions. Even small deviations from coplanarity (e.g. mutual inclinations $\sim 3^\circ$) could stir peak-to-peak obliquity variations up to $\sim 20^\circ$. Undetected planetary companions and/or the existence of a satellite could also destabilize the low-obliquity regions. In all cases, the high-obliquity region allows for moderate variations, and all obliquities corresponding to retrograde motion (i.e. $> 90^\circ$) are stable.

\bigskip


\end{abstract}


\section{Introduction}
The rapidly growing arsenal of exoplanet detections has greatly improved our understanding on the occurrence and orbital and structural properties of planetary systems \citep[e.g.,][]{Lissauer14, Winn15}. In particular, the NASA {\it Kepler} mission has discovered thousands of planetary candidates and identified terrestrial planets in the habitable zone (HZ) of their host stars \citep[e.g.,][]{Barclay13, Borucki13, Quintana14, Torres15}. The HZ is conventionally defined as the region where liquid water may exist on the surface of the planets with atmospheres similar to that of the Earth \citep{Kasting93, Kopparapu13}. The {\it Kepler} Habitable Zone Working Group has provided a list of HZ planets, based on various HZ boundaries and planetary radii, which includes 104 planetary candidates in the optimistic HZ and 20 planets with radii less than 2 $R_{\oplus}$ in the conservative HZ \citep{Kane16}. Many of such systems contain multiple planets, and their dynamical interactions could play a critical role in determining the habitability of these systems. \citet{Kane16} analyzed the dynamical stability of the potentially habitable multiplanetary systems, which also serves to validate the planetary systems. 

In addition to orbital stability, planetary obliquity and its variations are also important considerations in the habitability of a planet. Obliquity (or axial tilt) measures the angle between a planet's spin and orbital axes. These values are known for planets in the solar system. However, no reliable values for exoplanets have been claimed to date. Methods to infer exoplanetary spin axis direction from light curves via the effect of rotational flattening on transit depth and infrared phase curves \citep[e.g.][]{Gaidos04,Carter10}, as well as from high-contrast direct imaging through observing seasonal variations \citep{Kane17}, have been proposed. With very high quality photometric data and sophisticated modeling, obliquity measurements may be possible for the most favorable exoplanets in the future. 

Obliquity determines the latitudinal distribution of solar radiation on a planet and affects the modulation of its climate \citep{Williams97, Chandler00, Jenkins00, Spiegel09}. According to the Milankovitch theory, ice ages on the Earth are closely associated with variations in insolation at high latitudes, which depends on the orbital eccentricity and orientation of the spin axis \citep{Hays76, Weertman76, Imbrie82, Berger92}. At present, the obliquity variation of the Earth is regular and only undergoes small oscillations between $22\fdg1$ and $24\fdg5$ with a 41,000 year period \citep{Vernekar72, Laskar93b}. This is not to say that obliquity instability is wholly incompatible with life -- based on a simple 1D energy-balance atmospheric model, \citet{Armstrong14} suggest that large and frequent obliquity variations could help maintain higher surface temperatures and extend the outer boundary of the traditional HZ. 

Evolution in a planet's obliquity is governed by orbital perturbations from its companion planets, as well as torque from the host star and any moons acting on the planetary spin axis \citep[e.g.,][]{Ward74,Laskar93a}. When the perturbing frequencies from the companion planets match with the precession frequency due to the torques, the obliquity variation amplitude will increase due to resonant interactions. For instance, without the Moon, the torquing frequency of the Sun would match that from the companion planets, and the obliquity variation of the hypothetical Earth would be large (though constrained between $0$ and  $45^\circ$) over billion-year timescales \citep{Laskar93a, Lissauer12, Li14}. With the presence of the Moon, the obliquity variation of the Earth is significantly suppressed \citep{Laskar93b}. Other notable examples of planets whose spin axis dynamics have been thoroughly studied include Venus and Mars. For an early Venus with less atmospheric tides, obliquity variation in the low-obliquity range is small \citep{Barnes16}. On the other hand, the obliquity of Mars can vary with large amplitudes \citep{Ward73, Laskar93b, Touma93}. The obliquity changes of Mars likely resulted in runaway condensation of $\ce{CO_2}$ in the atmosphere, rendering Mars uninhabitable \citep{Toon80, Fanale82, Pollack82, Francois90, Nakamura03, Soto12}.

Prior to the advent of {\emph{Kepler}}, \citet{Atobe04} studied the obliquity variability of hypothetical HZ terrestrial planets co-inhabiting systems with giant planets. They characterized variation amplitudes as a function of rotation period and orbital distance for specific planet configurations. These include hypothetical systems with up to two giant planets with assorted masses and orbital elements, as well as real systems with known RV planets. Most known systems at the time were giant planets around sun-like stars. Though they did not explicitly consider compact systems with multiple small planets, \citet{Atobe04} arrived at many important general realizations to which we will refer in this article.  

Now, {\emph{Kepler}} has provided us with concrete systems to study, the kind with real, potentially habitable terrestrial planets neighboring multiple other small planets. Here, we will consider the obliquity variation of Kepler-62f and Kepler-186f, which are likely terrestrial planets in the HZ orbiting around a K2V-type star and an M1-type star, respectively \citep{Borucki13,  Quintana14}. Both of these two planets stay far away from their host stars, where the tidal influences of their host stars are comparatively weak. In particular, \citet{Bolmont14, Bolmont15} and \citet{Shields16} demonstrated that, for reasonable assumptions of the planetary properties and system ages, it is possible that Kepler-62f and Kepler-186f have not yet evolved to be tidally locked. This would allow them to keep high obliquities and short rotation periods. Note that the tidal evolution of planetary spin axis for general configurations shows sensitive dependence on the planetary tidal $Q$-values, as well as other assumptions (for more discussion, see Section \ref{ss:tides} and \citet{Heller11}). Using the Community Climate System Model, \citet{Shields16} identified combinations of orbital and atmospheric properties that permit surface liquid water for Kepler-62f, exploring both low- and high-obliquity regimes. To evaluate the potential habitability of these planets, these pioneering works considered many relevant factors, including long-term spin evolution. However, the effect of planetary spin-orbit coupling has not been investigated in detail. In addition, the mutual inclination between the planets has been assumed to take the minimum values (i.e. taken directly from the line-of-sight inclination of these planets), which leads to small obliquity variations. In reality, a larger range of inclinations is permitted by observation. Existing studies are also confined by their reliance on expensive {\emph N}-body simulations, which limits the set of parameter values subject to exploration.

In this article, we focus on the spin axis variability of the HZ planets on shorter timescales. We start by assuming that all planets in the system have been detected and study the evolution of the two five-planet systems. We relax the mutual inclination assumption and include a wide range of planetary system parameters consistent with observational constraints. We present a secular analytical framework applicable in the situation of small orbital eccentricities and inclinations. Such a framework is powerful because it allows us to visualize and predict the nature of obliquity variations in a large parameter space, as well as examine the sensitivity of conclusions to errors in the observed parameters. Specifically, we consider different planetary rotation rates and additional planets and satellites, and we characterize regions in this parameter space where the resonant interactions between the HZ planet and its companions may cause large obliquity variations. We briefly discuss the prospects of long-term obliquity evolution subjected to the gradual but inevitable tidal synchronization process.

This paper is organized as follows: planetary system properties used throughout the paper are explained in section \textsection \ref{s:sys-pars}. In section \textsection \ref{s:nr}, we use {\emph N}-body simulations coupled with secular integration to illustrate the evolution of obliquity. We consider variations in obliquity as a product of resonant interactions using an analytical approach and interpret our numerical results in section \textsection \ref{s:ar}. Section \textsection \ref{s:dis} explores the effects of undetected planets, satellites, and the possible path to tidal synchronization. A summary is presented in section \textsection \ref{s:conclusion}.

\bigskip

\section{Planetary System Parameters}\label{s:sys-pars}

We anchor our analysis on the measured properties of the two {\emph{Kepler}} systems with potentially habitable planets and infer the rest of the relevant parameters under assumptions discussed below. Parameters and their representative values used in this work are given in Table \ref{t:sys-pars}. 

\input{System_pars.tex}

Most direct observables in exoplanet systems are combinations of properties of the planet and the host star. Therefore, any indeterminacy in the stellar properties is directly propagated into errors in the planet properties. For instance, the radius of a planet and its semi-major axis are derived from the transit depth and orbital period in concert with the stellar radius ($R_\star$) and mass ($M_\star$), respectively. We set the mass and radius of Kepler-62 to be $M_\star = 0.69\pm0.02 M_\sun$ and $R_\star = 0.63\pm0.02 R_\sun$ \citep{Borucki13}, and that of Kepler-186 to be $M_\star = 0.544~\pm0.02 M_\sun$ and $R_\star = 0.523~\pm0.02 R_\sun$ \citep{Torres15}. Measurements of orbital periods ($P_{\rm orb}$) from the transiting technique tend to be highly accurate. The semi-major axes ($a$) can in turn be calculated from $M_\star$ and $P_{\rm orb}$ using Kepler's third law. We adopt the semi-major axes for the Kepler-62 planets from \citet{Borucki13}, and we calculate the semi-major axes for Kepler-186 based on the updated stellar mass from \citet{Torres15} and the orbital period measurements from \citet{Quintana14}. The age of the Kepler-62 system is determined to be $7 \pm 4$ Gyr \citep{Borucki13}, though recently \citet{Morton16} arrived at a much younger age of $2.34^{+2.15}_{-1.02}$ Gyr. Kepler-186 is estimated to have an age of $4 \pm 0.6$ Gyr \citep{Quintana14}. 

The line-of-sight orbital inclinations ($i_{\rm LoS}$) are computed from the measured impact parameters ($b$), as well as the stellar radius ($R_\star$) and semi-major axes: 
\begin{equation}
i_{\rm LoS} = \arccos\left[\frac{b R_\star}{a}\right].
\label{e:inclination}
\end{equation} 
We obtain the $i_{\rm LoS}$ for planets in Kepler-62 from \citet{Borucki13} directly, and we calculate those for Kepler-186 based on the measurements of the $b$ values in \citet{Quintana14} and the updated stellar parameters from \citet{Torres15}. Although the line-of-sight inclinations may not directly translate into the inclination of the planets measured from a reference plane, the fact that the planets transit means that they are most likely arranged in nearly coplanar configurations. We use $90^\circ - i_{\rm LoS}$ as the minimum initial orbital inclinations in our numerical study, consistent with previous works \citep[e.g.][]{Bolmont14, Shields16}. While the eccentricities are not well constrained, they are all reasonably close to 0. For simplicity, we assume that all orbits are initially circular in the numerical simulations. 

Since the planetary masses ($M_p$) have not been directly measured for these systems, we use the publicly available code {\tt{Forecaster}} of \citet{ChenKip17} based on a probabilistic approach that predicts masses for a variety of celestial bodies from their radii. We directly use the published planetary radii ($R_p$) and their errors as given in Table 1 of \citet{Borucki13} for Kepler-62, and we calculate $R_p$ using transit depth measurements from \citet{Quintana14} and stellar parameters from \citet{Torres15}, propagating both errors for Kepler-186. For a given input radius, the output is a probability distribution of the possible masses, where {\tt{Forecaster}} has marginalized over a range of possible planetary compositions. The resultant uncertainty in the forecasted masses can be quite large and asymmetric (on order of $\sim 30 - 70\%$). While the measured radii for Kepler-62f ($1.41 \pm 0.07 R_\earth$) and Kepler-186f ($1.17 \pm 0.11 R_\earth$) are consistent with planets with rocky compositions ($< 1.62 R_\earth$, \citet{Rogers15}), in its standard implementation {\tt{Forecaster}} also considers Neptunian compositions. Since we study the habitability of Kepler-62f and Kepler-186f in this work, we presuppose that they are rocky. Thus, to enforce the condition that they be terrestrial planets, we inverted the Terran mass-radius power law given in \citet{ChenKip17} to generate the default masses for Kepler-62f and Kepler-186f used in this work: 

\begin{equation}
\left(\frac{M_p}{M_\earth}\right) = \left[\frac{1}{1.008} \left(\frac{R_p}{R_\earth}\right)\right]^{1/0.279}. 
\end{equation}

\noindent The resulting masses are 3.3 and 1.7 $M_\earth$ for Kepler-62f and Kepler-186f, respectively, similar to those considered in existing studies \citep{Bolmont14, Bolmont15, Shields16, Quarlesinprep}. We feed the median radii into either {\tt{Forecaster}} (all planets except 62f and 186f) or the Terran power-law relation, where we marginalize over hyperparameter posteriors given in \citet{ChenKip17} (62f and 186f), and tabulate the default and 68\% symmetric confidence intervals for the planetary mass posteriors in Table \ref{t:sys-pars}.  


\bigskip
\section{The Evolution of Obliquity over Time: Numerical Results}
\label{s:nr}

\begin{figure}[htb]
\includegraphics[scale=0.6]{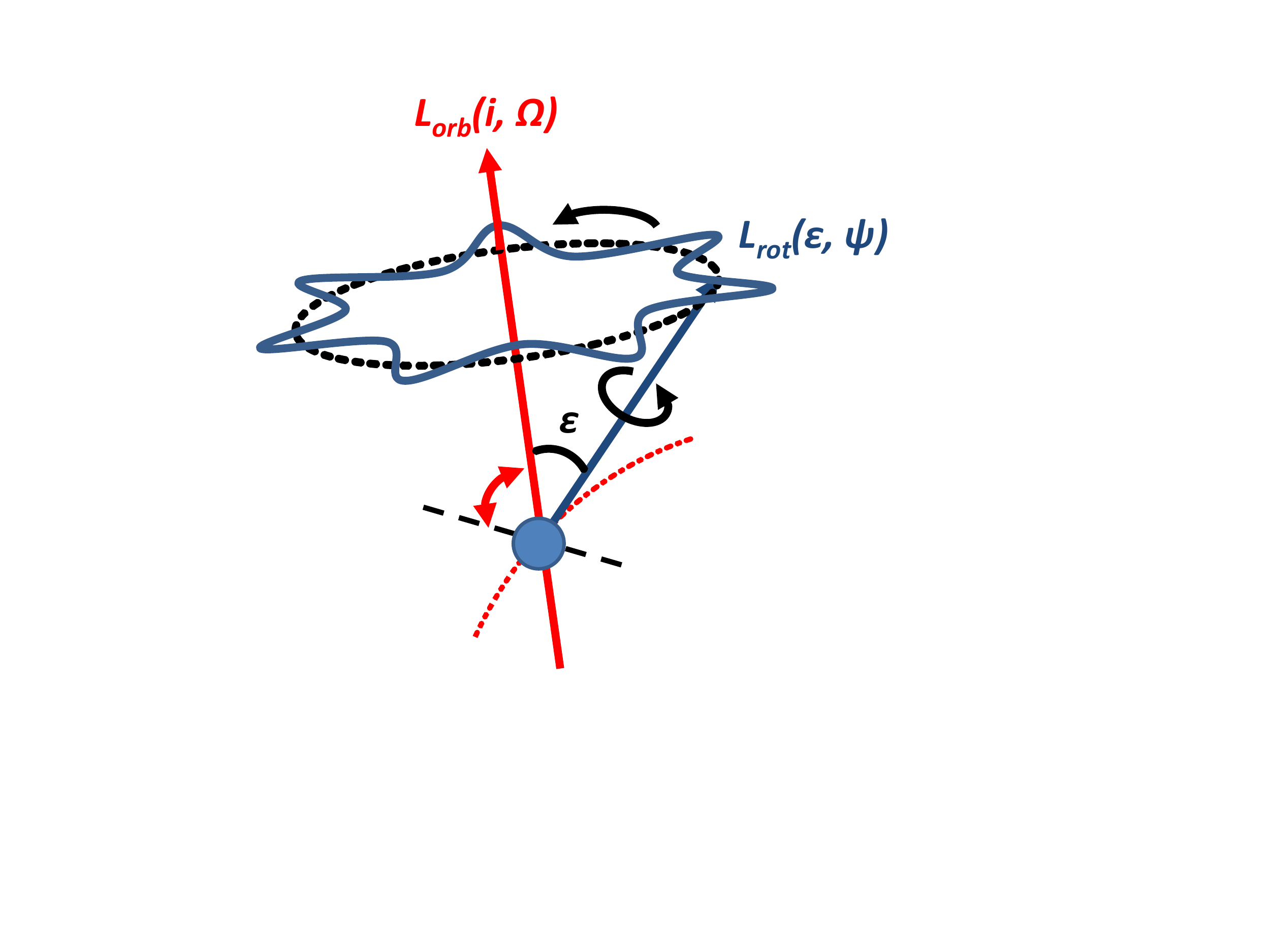}
\caption{Geometry of obliquity evolution. This schematic diagram depicts the orbital ($L_{\rm orb}$) and spin ($L_{\rm rot}$) axes and the angles involved. The combined evolution of these two angles forms the trajectory of precession and nutation. $L_{\rm orb}$ can be torqued by other bodies in the system, which could interact resonantly with $L_{\rm rot}$ and cause it to oscillate. See also Fig 1 in \citealt{Atobe04}.}
\label{f:cartoon}
\end{figure}

An illustration of the planetary spin-orbit misalignment (obliquity) is shown in Figure \ref{f:cartoon}. $L_{\rm orb}$ and $L_{\rm rot}$ denote the angular momentum vector of the orbit and the planet, respectively, and the obliquity angle ($\epsilon$) represents the angle between $L_{\rm orb}$ and $L_{\rm rot}$. The inclination $i$ and the longitude of node $\Omega$ characterize the orientation of the orbital plane, or the direction of $L_{\rm orb}$. Similarly, the obliquity angle $\epsilon$ and the longitude of the spin axis $\psi$ determine the orientation of the planetary spin axis with respect to the orbital plane.

The phenomenon of planetary obliquity variations is induced by the torques of the the host star on the equatorial bulge of the planet, as well as by periodic forcing from other planets. Therefore, its dynamics depends on the orbital configuration of all the bodies in the system. In this section, we use a numerical approach to study short-term obliquity evolution. We present the framework for the numerical studies in section \textsection \ref{ss:ham}, and numerical results on the obliquity variation of a few representative manifestations for the two five-planet systems are shown in \textsection \ref{ss:eps-t}. A qualitative comparison to existing spin-dynamical studies on these systems is given in \textsection \ref{ss:compare}. 

\bigskip



\bigskip
\subsection{Numerical Method Framework}\label{ss:ham}

To calculate the obliquity evolution of a planet, we first perform {\emph N}-body simulations to obtain its orbital evolution. Then, we compute the spin axis dynamics, which is coupled to the planetary orbit, based on the {\emph N}-body results and the Hamiltonian below (Eqn.(\ref{e:ham})). 
This approach implicitly assumes that the planets' orbital angular momenta are dominant over that of their spin, which allows us to ignore the feedback of the spin axis to the planetary orbit.  

We use the hybrid integrator in the publicly available {\tt{Mercury}} {\emph N}-body code \citep{Chambers99} to simulate representative manifestations for each planet system for $10^7$ yr, stepping in increments of 0.5 days. The initial values of semi-major axes, eccentricities, and mutual inclinations are discussed in Section \ref{s:sys-pars}. 

The spin-orbit coupling is described by the secular Hamiltonian of obliquity variation, which is well documented in the literature \citep[e.g.][]{Goldreich66,Wisdom84,Laskar93a,deSurgy97}: 

\begin{align}
H(\chi,\psi,t) = &  \frac{1}{2}\alpha_p\chi^2+\sqrt{1-\chi^2}  \nonumber \\
&  \times [A(t) \sin \psi + B(t) \cos \psi].
\label{e:ham}
\end{align}

\noindent Here $\chi$ and $\psi$ are the Andoyer canonical variables, where $\chi=\cos{\epsilon}$ and $\psi$ is the longitude of the planet's spin axis \citep{Andoyer23, Kinoshita72}. $\epsilon$ is the obliquity, and $A(t)$ and $B(t)$ are functions of $p = \sin(i/2)\sin\Omega$ and $q=\sin(i/2)\cos\Omega$, where $i(t)$ represents orbital inclination and $\Omega(t)$ is the longitude of the ascending node of the planet:

\begin{equation}
A(t) = 2(\dot{q}+p(q\dot{p}-p\dot{q}))/\sqrt{1-p^2-q^2},
\label{e:at}
\end{equation}
\begin{equation}
B(t) = 2(\dot{p}+q(p\dot{q}-q\dot{p}))/\sqrt{1-p^2-q^2}.
\label{e:bt}
\end{equation}

$\alpha_p$ is the precession coefficient, defined for a given planet $p$ as

\begin{equation}
\label{e:alpha}
\alpha_p =  \frac{3}{2\omega}\left[\frac{GM_\star}{(a_p\sqrt{1-e_p^2})^3}\right]E_d ,  
\end{equation}
where $a_p$ and $e_p$ denote the semi-major axis and eccentricity of the planetary orbit around the star, respectively. $\omega$ is the angular velocity of the planet's spin. $E_d = (C-1/2(A+B))/C$ is the dynamical ellipticity, where $A$, $B$ and $C$ are the moment of inertia along the three principal axes. $E_d$ is related to the oblateness (flattening) of the planet and generally scales with $\omega^2$ \citep{Lambeck80}. For a moonless Earth (i.e. only considering the torque from the Sun), $\alpha_\Earth = 17\farcs4 ~{\rm yr}^{-1}$ \citep{Laskar93a}.

As the internal structure and the dynamical ellipticity of the exoplanets are unknown, we assume that the dynamical ellipticity is the same as that of Earth if the planet rotates with the same period as Earth's.\footnote{With very high precision light curves, it may become possible to measure exoplanetary oblateness, which is related to $E_d$, from transit depth variations \citep{Carter10,Biersteker17} and ingress/egress anomalies \citep{Zhu14}, though it would be very difficult in general.} For planets on nearly circular orbits ($e_p \sim 0$), the relation for $\alpha_p$ is then given as

\begin{equation}
\label{e:alpha_earth}
\alpha_p = \left(\frac{P_{\rm rot}}{\rm day}\right)^{-1}\left(\frac{M_\star}{M_\sun}\right) \left(\frac{a_p}{\rm au}\right)^{-3} \alpha_\Earth, ~~ \alpha_\Earth = 17\farcs4 ~{\rm yr}^{-1}. 
\end{equation}

\noindent Note $\omega \propto P_{\rm rot}^{-1}$. Scaling the precession coefficient with the host star mass and the semi-major axis of each planet, Eqn.(\ref{e:alpha_earth}) gives $\alpha_{K62f} = 32\farcs2~{\rm yr}^{-1}$ and $\alpha_{K186f} = 137\farcs4~{\rm yr}^{-1}$ corresponding to planetary rotation periods of 24 hr. This approach yields similar results to that of \citet{Lissauer12} \citep[see also][]{Quarlesinprep},  who assumed that the moment of inertia coefficient ($C/(M_pR_{eq}^2)$) of the planet is the same as Earth's, where $M_p$ is the planetary mass and $R_{eq}$ is the equatorial radius of the planet. In particular, for the planetary masses and radii assumed in Table \ref{t:sys-pars} and rotation periods of 24 hours, $\alpha_{K62f} = 31\farcs9~{\rm yr}^{-1}$ and $\alpha_{186f} = 134\farcs4~{\rm yr}^{-1}$ based on the approach by \citet{Lissauer12}. We set $\alpha_{K62f} = 32\farcs2~{\rm yr}^{-1}$ and $\alpha_{K186f} = 137\farcs4~{\rm yr}^{-1}$ as our default, Earth-like values. Of course, the rotation state of an arbitrary planet may not resemble that of the Earth. Planets formed from protoplanet accretion in {\emph N}-body simulations tend to have rapid primordial spins \citep[e.g.][]{Kokubo07}. Over time, $P_{\rm rot}$ is modified by tidal interaction with the host star and any satellites present. To explore rotation periods deviating from that of the Earth, we scale the $\alpha_{\rm{p}}$ values with the inverse of the planet's rotation period ($P_{\rm rot}$), as in Eqn.(\ref{e:alpha_earth}). 


\bigskip
\subsection{Obliquity Evolution for Variable Rotation Periods, and Orbital Inclinations}\label{ss:eps-t}

Theoretically, a planet's initial obliquity is often determined by stochastic impacts during formation and can be expected to take an isotropic distribution \citep[e.g.][]{Kokubo07}. Therefore, we explore the dynamical evolution of the obliquity starting with a wide range of initial values, $\epsilon_0$, \footnote{In all our integrations, we started with initial longitude of spin axis at $0^\circ$ (i.e. $\psi_0 = 0^\circ$). The amplitude of variability has some dependence on $\psi_0$ \citep[e.g.,][]{Atobe04, Quarlesinprep}. Different $\psi_0$ can produce variation amplitudes that differ by a factor of up to $\sim 3$.}, from prograde ($\epsilon < 90^\circ$) to retrograde ($\epsilon > 90^\circ$). Hereafter, references to `low' and `high obliquity' are relative to the prograde regime only.

\begin{figure*}[htb]
\begin{minipage}{0.5\textwidth}
\centering
\includegraphics[width=3.6in]{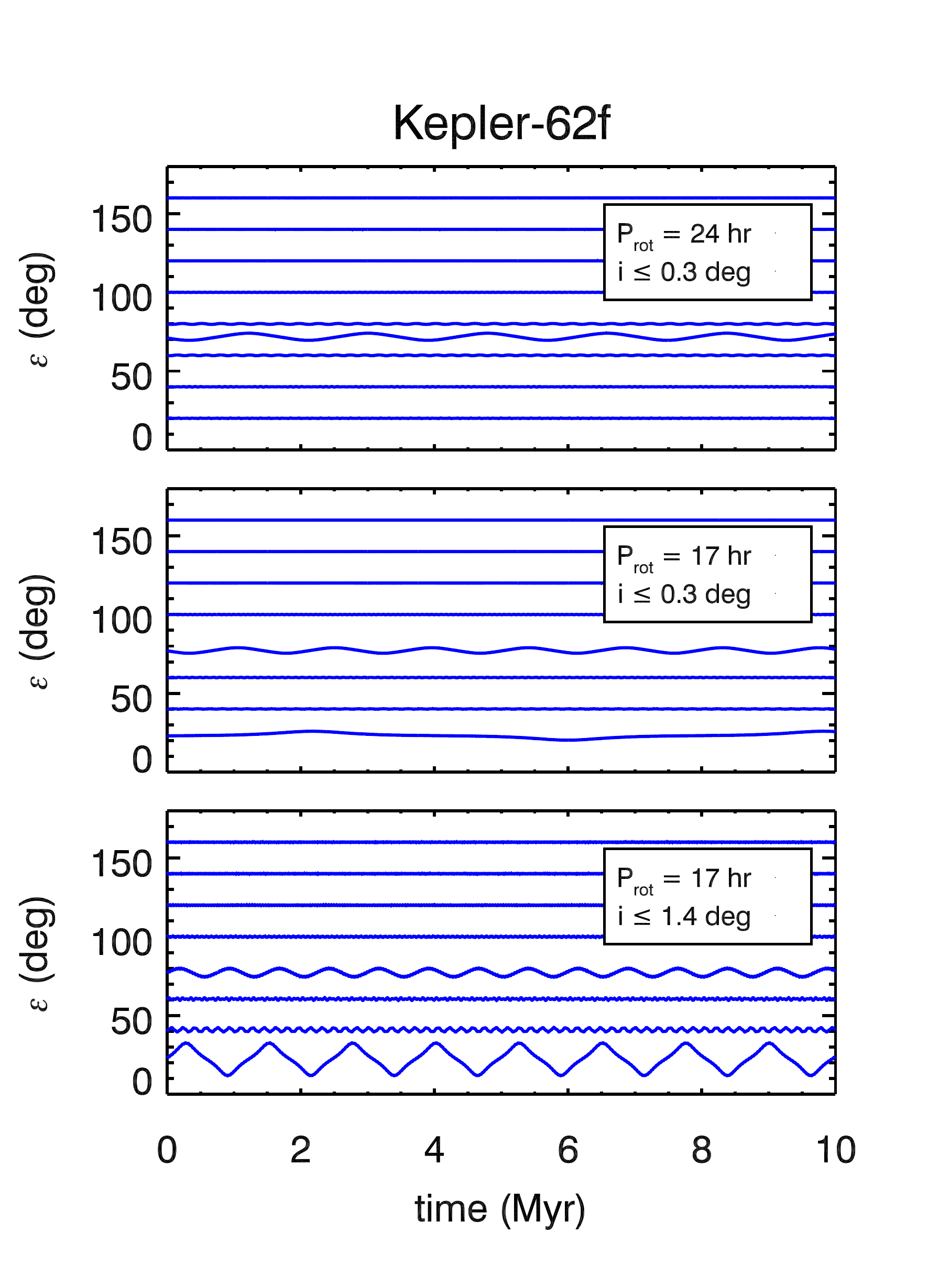}
\end{minipage}%
\begin{minipage}{0.5\textwidth}
\centering
\includegraphics[width=3.6in]{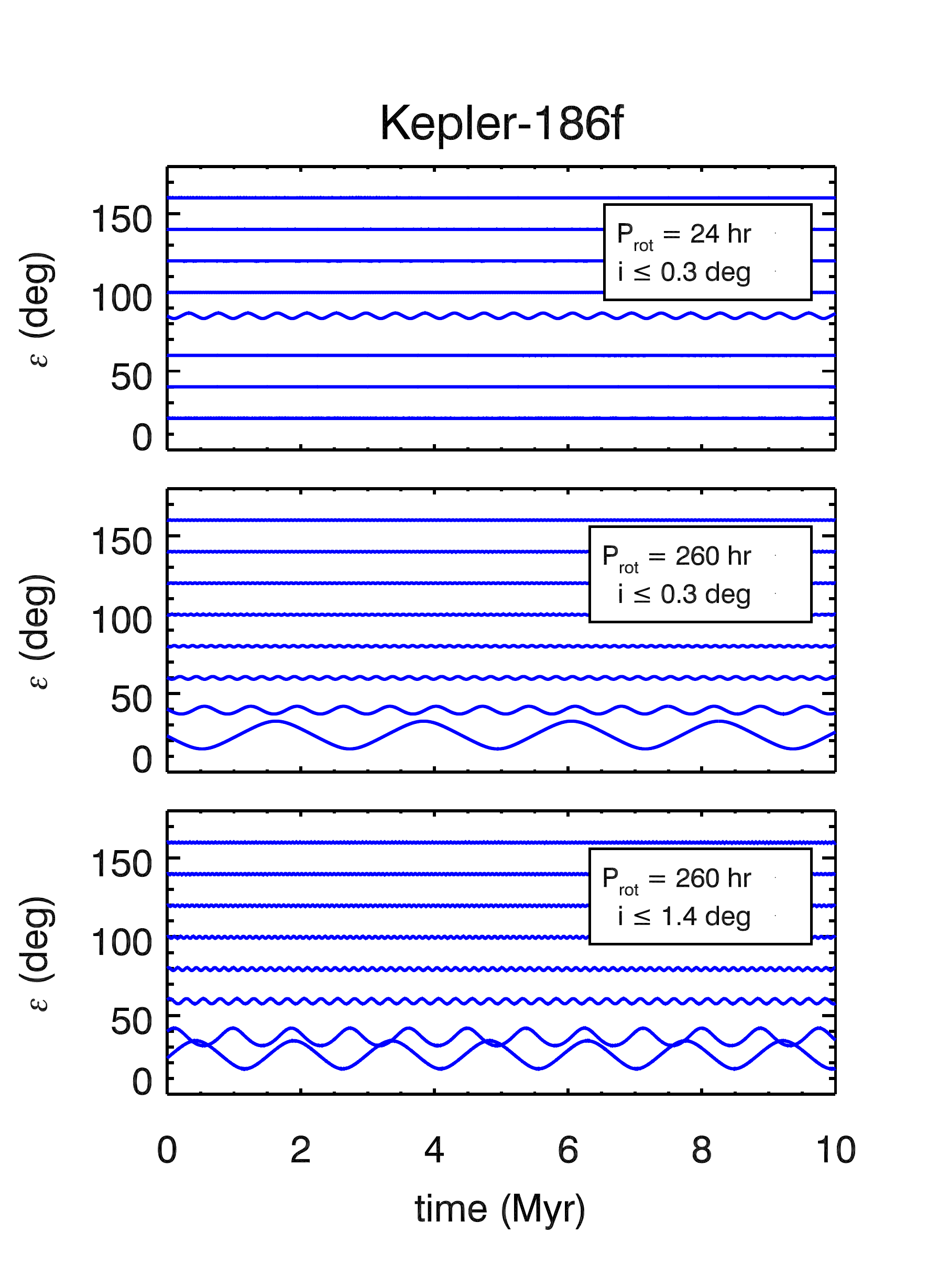}
\end{minipage}
\caption{Evolution of obliquity over $10^7$ yr for representative initial obliquity values ($\epsilon_0$). Left panels: Kepler-62f. Right panels: Kepler-186f. The top panels correspond to a planetary rotation period of 24 hr. For both planets, the time series at the Earth-like obliquity of $20^\circ$ have very small variabilities, which are compatible with results from \citet{Bolmont14} (Kepler-186f) and \citet{Bolmont15} (Kepler-62f). The middle and bottom panels display the behavior of obliquity at select non-Earth-like planetary rotation periods. In particular, $P_{\rm rot}$ is 17 hr for Kepler-62f and 260 hr for Kepler-186f. The lowest panels correspond to the configuration with slightly larger planet-planet mutual inclinations ($\lesssim 3^\circ$), where the maximum orbital inclinations of Kepler-62f and Kepler-186f reach $\sim1\fdg5$. The amplitude of variability is generally increased with orbital inclination. In all cases, obliquities corresponding to retrograde spin (i.e. $\epsilon > 90^\circ$) are stable.}
\label{f:eps-t}
\end{figure*}

For particular realizations of the Kepler-62 and Kepler-186 systems, the evolution of the obliquity angle is shown in Figure \ref{f:eps-t}. The left and right columns correspond to the results for Kepler-62f and Kepler-186f, respectively. The top panels assume that the planetary rotation period is 24 hr, similar to our Earth. In this case, the low-obliquity regions have low variability for both Kepler-62f and Kepler-186f, while the high-obliquity regions allow small variabilities for Kepler-62f. The timescale of oscillation ranges from a fraction of to several megayears. In both cases, retrograde obliquities show little variation.

For particular values of the planetary rotation period  (i.e. proxy for $\alpha_p$), the lower region of the obliquity allows larger variabilities. Examples of such rotation periods are approximately $17$ hours ($\alpha \approx 45\farcs5~{\rm yr}^{-1}$) for Kepler-62f and $260$ hr for Kepler-186f ($\alpha = 12\farcs7~{\rm yr}^{-1}$). The resultant obliquity evolutions are shown in the middle panels of Figure \ref{f:eps-t}. We will illustrate in section \textsection \ref{s:ar} that the locations and natures of obliquity variation are determined by the properties of resonant interactions. 

In the bottom panels, we consider slightly higher inclinations, requiring the mutual inclinations of the planets to be within $3^\circ$, in accordance with studies on the mutual inclination of multiplanetary systems \citep{Fang12,Fabrycky14,Ballard16,Moriarty16}. We set the initial longitude of ascending nodes of the planets such that all of the planets transit over 4 yr timescales, i.e. over the tenure of the original {\emph{Kepler}} mission. The maximum inclinations of Kepler-62f and Kepler-186f reach $\sim 1\fdg5$ over 10 Myr. We adopt the same precession coefficient (i.e. assume the same rotation periods and interior mass distribution) as those in the middle panels. With the slightly higher mutual inclinations, the perturbation to the planetary orbit is stronger, leading to enhanced obliquity variation (up to $\sim 20^\circ$ in peak-to-peak amplitude). We obtain an analytical expression for the amplitude of variabilities in section \textsection \ref{ss:res-wid}. 

Obliquity is affected by both the orbital inclination and the orientation of the spin axis relative to a fixed reference plane. Naturally, the intrinsic variabilities of the two factors both contribute to the overall observed variability in the obliquity. In regions where the strongest variabilities occur (due to resonant spin-orbit coupling; see \textsection 4), the variation is dominated by that of the spin axis. Weaker variability can manifest in other regions (due to higher-order effects; \textsection 4), particularly visible in the bottom panels of Figure \ref{f:eps-t}. In these cases, oscillation of the orbital inclination and spin axis orientation both contribute comparably to the obliquity variations.

%

\bigskip
\subsection{Comparison to Previous Studies}\label{ss:compare}

The obliquity evolutions for these two exoplanets were previously explored, albeit usually for either a narrow range of initial conditions and/or for a much longer term under tidal influence. Directly comparable with our study are the 10 Myr results, in which \citet{Bolmont14} find that Kepler-186f's obliquity is stable if it starts with an Earth-like obliquity ($23^\circ$) and rotation period ($\sim$ 24 hr). \citet{Bolmont15} reach the same conclusion for Kepler-62f. These are very consistent with the outcomes found in this work for the same set of assumptions, as shown in the top panels of Figure \ref{f:eps-t}. However, more complex behavior is possible with different $\epsilon_0$, rotation periods, and mutual inclinations, as illustrated in the bottom panels in this figure. 

During the final preparation of our manuscript, we noticed that \citet{Quarlesinprep} have been working on a similar problem related to the obliquity variation of Kepler-62f, using a different approach. Specifically, adopting direct {\emph N}-body simulations including spin-orbit coupling, \citet{Quarlesinprep} find that the low-obliquity region of Kepler-62f is stable, assuming that it is an Earth analog. This is consistent with our results. 

\bigskip

\section{The Theory of Obliquity Instablity and Results from an Analytical Framework}
\label{s:ar}

The resonant interactions between the torque from the host star acting on the planetary spin axis and the orbital perturbation from the companion planets can be quantified in a straightforward, analytical framework. This can in turn be used to rapidly approximate the obliquity evolution for given properties of the orbital architecture and that of the planet of interest. In this section, we summarize the analytical approach to characterize the locations of the resonant regions (section \textsection \ref{ss:res-zones}), and we derive an analytical expression for the size of the resonant region  for Kepler-62f and Kepler-186f  (section \textsection \ref{ss:res-wid}). Then, we interpret the numerical results based on the analytical approach (section \textsection \ref{ss:num-freq}). In the end, we characterize the parameter space of obliquity evolution considering the uncertainties in the observational orbital parameters and planetary masses based on the analytical approach (section \textsection \ref{ss:ll-th}). The basis of this analysis lies in the simplified Hamiltonian in Eqn.(\ref{e:ham}), which is analogous to that of a physical pendulum.  


\bigskip
\subsection{Location of Resonances}\label{ss:res-zones}

The expected resonant locations can be calculated analytically by identifying the obliquity values that allow resonant coupling between the precession of the planetary spin axis and the oscillation of the planetary orbital plane, as discussed extensively in the literacture \citep[e.g.,][]{Laskar93a, Touma93, Lissauer12}. Specifically, the form of Eqn.(\ref{e:ham}) is analogous to the Hamiltonian describing a physical pendulum with angular position $\theta$ and angular momentum $\mathcal{L}$:

\begin{equation}
H_{\rm pendulum}(\mathcal{L},\theta)=\frac{\beta}{2}\mathcal{L}^2 + c\cos \theta.
\label{e:pen}
\end{equation}  

\noindent The $\beta$ term here is akin to $\alpha_p$ in Eqn.(\ref{e:ham}), while $c$ can be compared with $\sqrt{1-\chi^2}A(t) \sim \sqrt{1-\chi^2}B(t)$, as $A(t)$ and $B(t)$ are expected to be on the same order.

A natural frequency for the pendulum system is given by the characteristic rate of variation for $\theta$:

\begin{equation}
\dot{\theta} = \frac{\partial H_{\rm pen}}{\partial \mathcal{L}} = \beta \mathcal{L}, 
\end{equation}

\noindent and resonance occurs when a perturbing angular frequency ($f$) coincides with this natural frequency. Comparing to Eqn.(\ref{e:ham}), we see that $\alpha_p \chi$ is analogous to the natural frequency of this system (indeed, it is the frequency of the axial precession). Therefore, for obliquity $\epsilon$ to vary, this condition amounts to requiring $-f = \alpha_p \chi = \alpha_p \cos(\epsilon)$. In general, there can be multiple forcing frequencies. We use $f_k$ to denote the $k$th modal frequency of the forcing terms $A(t)$ and $B(t)$. Since $A(t)$ and $B(t)$ are functions of orbital inclination and the longitude of the ascending node, their characteristic frequencies should follow that of the inclination vector. 

Therefore, the $k$th resonance occurs at obliquity angle $\epsilon_{\rm{res}, k}$, where

\begin{equation}
\cos(\epsilon_{{\rm res}, k}) = -f_k/\alpha_p.
\label{e:coseps}
\end{equation}

\noindent It follows that negative frequencies correspond to obliquity resonances in the prograde regime ($\epsilon < 90^\circ$) and positive frequencies to retrograde ones ($\epsilon > 90^\circ$). Of course, only modes with frequency $f_k/\alpha_p \in [-1,1]$ lead to physical values of $\epsilon_{{\rm res}, k}$, and can result in resonant interactions. This equation can be inverted to read $\alpha_p = -f_k/\cos(\epsilon_{{\rm res}, k})$. Since, given $f_k$, there is a corresponding $\alpha_p$-value for every arbitrary $\cos(\epsilon_{\rm res})$,  we conclude that, for every obliquity value, each mode can induce instability at one fine-tuned rotation rate. This will become visually apparent in section \textsection \ref{ss:ll-th}.

\bigskip 
\subsection{Resonant Widths}\label{ss:res-wid}
Resonances are not point-like -- each has a finite extent generally centered around $\epsilon_{{\rm res}, k}$. The width of the resonance determines the variability amplitude of the obliquity. Although the actual variability amplitude may not span the full width of the associated resonance owing to a dependence on the initial longitude of the spin axis, $\psi_0$ (see also \textsection\ref{ss:eps-t}), resonance widths can serve as order-of-magnitude proxies to guide expectations. In this section, we calculate the resonant widths analytically and use them to characterize the amplitude of variability.  

To derive the width of the resonant zone, we again invoke the similarity between obliquity dynamics and that of the physical pendulum. For the physical pendulum (Eqn.\ref{e:pen}), the half-width of a resonant zone in $\mathcal{L}$ is $\Delta \sim 2\sqrt{c/\beta} \sim 2 (\sqrt{1-\chi^2}A(t)/\alpha_p)^{1/2}$ \citep[e.g.][]{Li14}. 

For the obliquity variation in a resonant region, $A(t)$ can be approximated as follows when the inclination is small:            

\begin{equation}
A(t)\sim \alpha_p\chi_{{\rm res},k} i_k.
\label{e:atsim}
\end{equation}

Thus, the half-width of the resonant region can be expressed as the following:
\begin{equation}
\Delta_{\chi, k} \sim 2 ( \chi_{{\rm res},k}\sqrt{1-\chi_{{\rm res},k}^2}  i_k)^{1/2},
\label{e:wid}
\end{equation}
which is identical to Eqn (11) in \citealt{Atobe04}, who derives this relation directly from the equation of motion for this problem. The expression for variability amplitudes at nonresonant obliquity values can be found in,e.g., \citet{Ward73}.

Note that in both Equations (\ref{e:atsim}) and (\ref{e:wid}) $i_k$ is measured in radians, and we assume $\chi$ to be near $\chi_{\rm res}$ in the resonant region. Thus, this approximation fails when the resonant zone is large ($\gtrsim 0.1 \chi_{\rm res}$). The width is independent of $\alpha_p$ because it is canceled out in the amplitude calculation. Note that Eqn.(\ref{e:wid}) does not apply when $\alpha = 0$, which corresponds to a completely rigid sphere with zero dynamical ellipticity. In this case, no precession is expected (and hence no resonance), and the obliquity variation follows that of the orbital inclination.

\begin{figure}[htb]
\hspace{-1cm}
\includegraphics[scale=0.55]{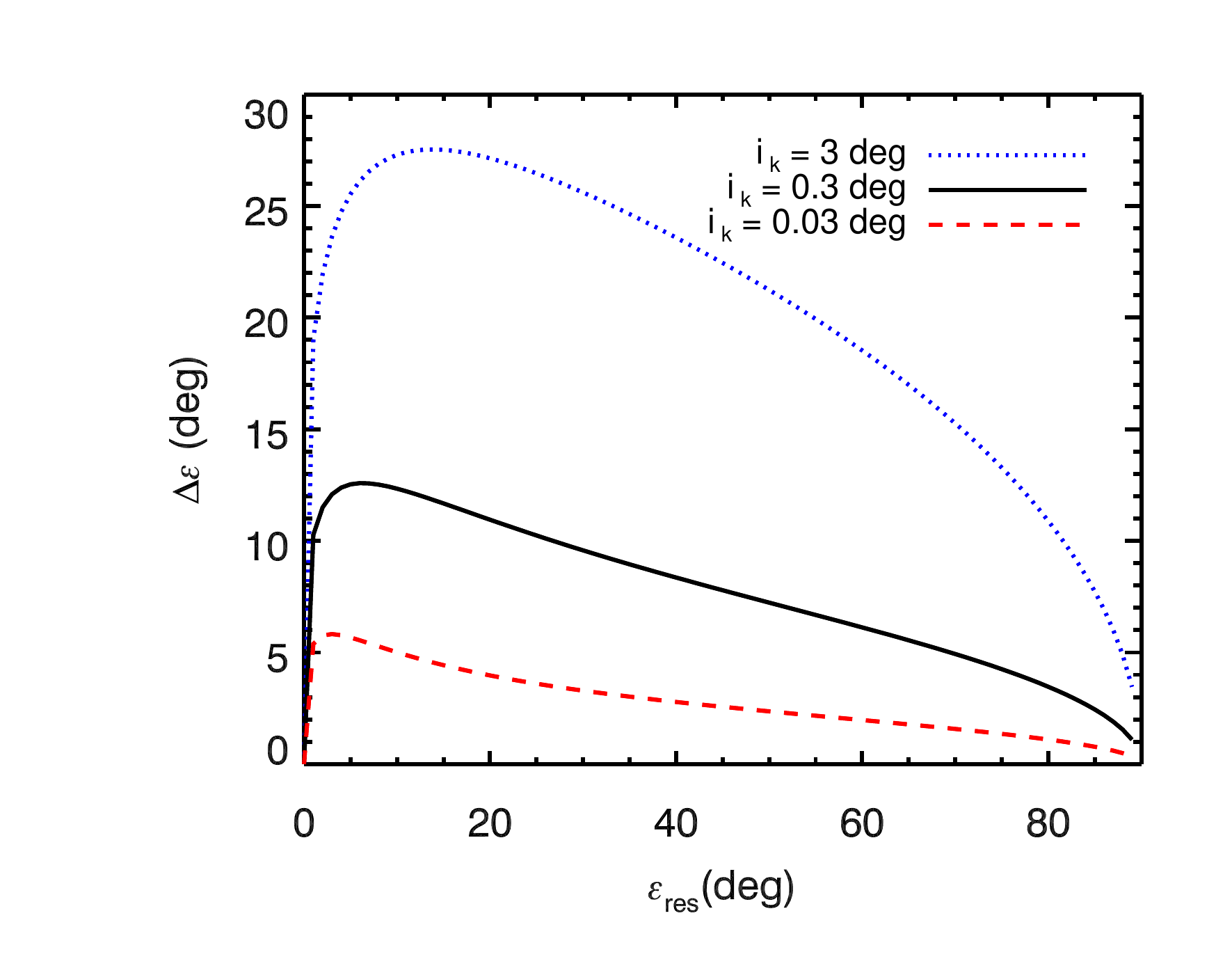}
\caption{Approximate resonant width in resonant zones ($\Delta \epsilon$) vs. the resonant obliquity ($\epsilon_{\rm res}$), following Eqn.(\ref{e:at}). Resonances tend to be wider at lower-obliquity values and for modes associated with larger inclination components. }
\label{f:res-widths}
\end{figure}

Figure \ref{f:res-widths} illustrates the half-width of the resonant zones as a function of the resonant obliquity values. The width is much larger in the low resonant obliquity region owing to the conversion from $\chi$ to $\epsilon$ being arc-cosine. The dashed, solid, and dotted lines correspond to mode amplitudes $i_k = 3^\circ$, $0\fdg3$, and $0\fdg03$, respectively. When the modal amplitude is small (e.g. $i_k \sim 0\fdg03$), the resonant width is also small. Therefore, the fact that a given obliquity value lies in the resonant region does not necessarily  imply that its variation amplitude must be large -- depending on the associated modal importance, the variability could very well be confined to within a few degrees. In general, planets whose orbits have a larger mutual inclination with a given planet will induce stronger modes on that planet's inclination. This mechanism is responsible for the enhanced variability amplitudes found in the numerical results for the more inclined cases in Figure \ref{f:eps-t}. 

\bigskip
\subsection{Modal Properties Based on {\emph N}-body Simulations and Predicted Obliquity Variations}\label{ss:num-freq}

\begin{figure}[htb]
\begin{minipage}{0.5\textwidth}
\centering
\hspace{-1cm}
\includegraphics[scale=0.5]{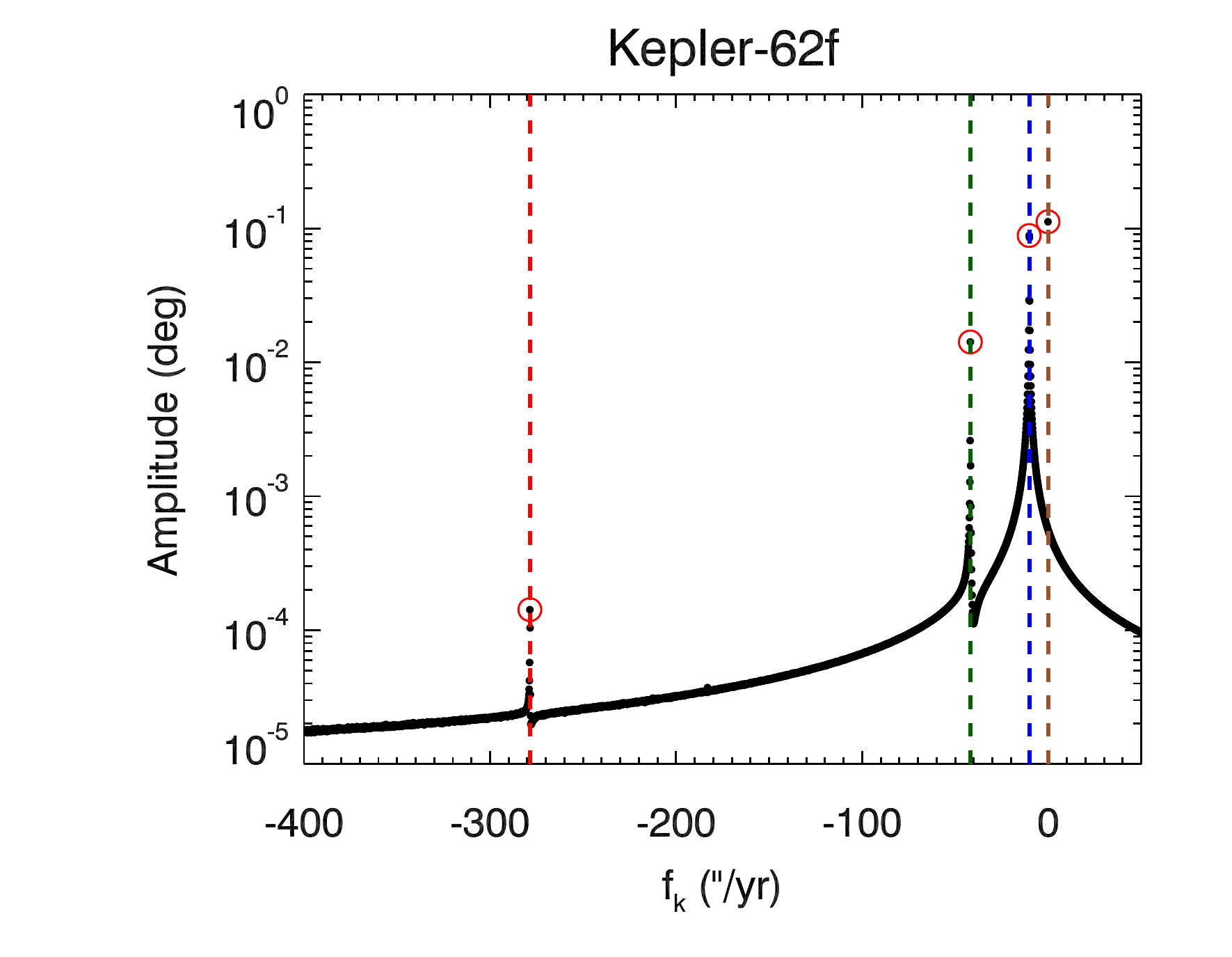}
\end{minipage}%
\begin{minipage}{0.5\textwidth}
\centering
\hspace{-1cm}
\includegraphics[scale=0.5]{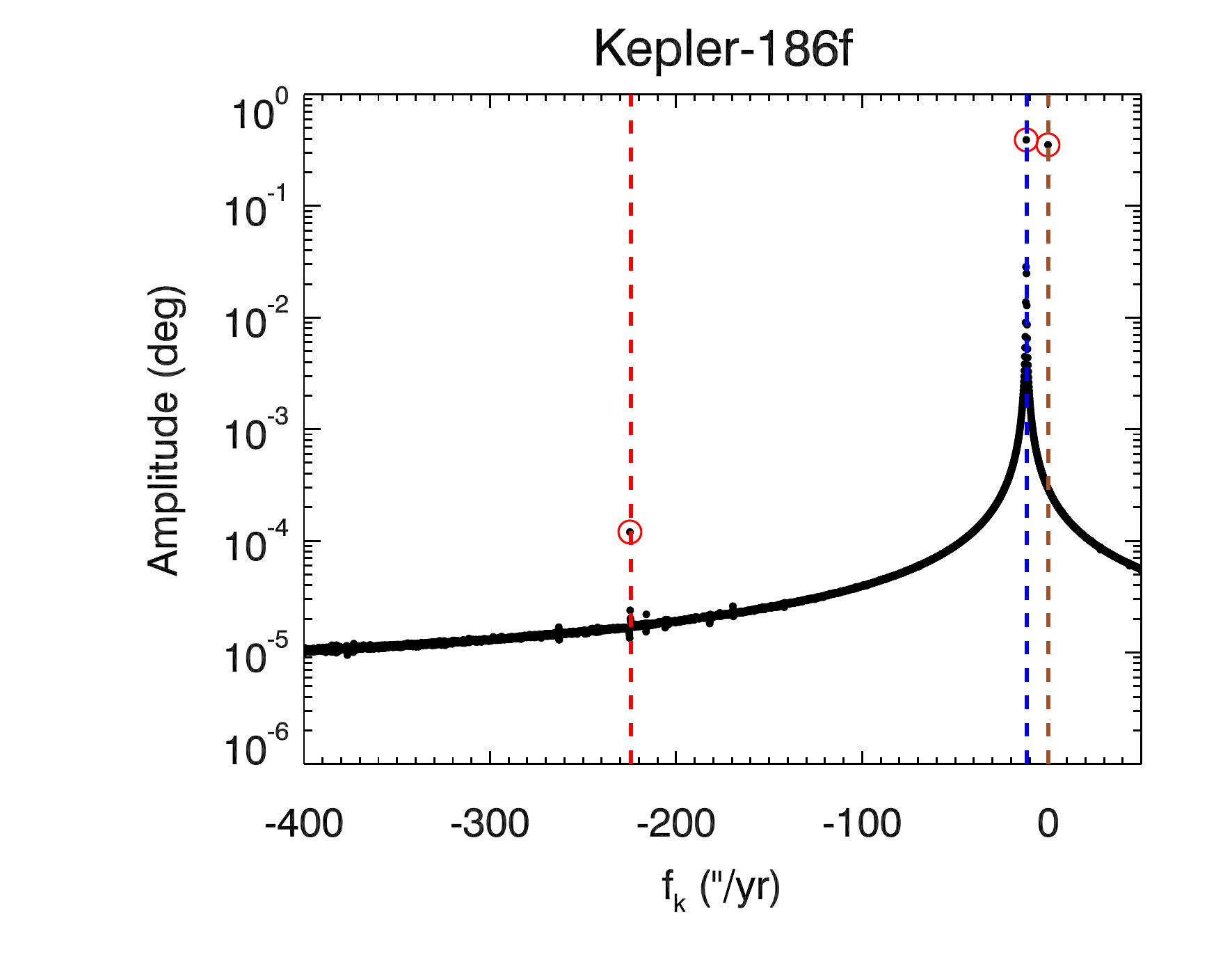}
\end{minipage}
\caption{Fourier transform spectrum of $i(t) e^{\sqrt{-1}\Omega(t)}$ from {\emph N}-body simulations for the near-coplanar case of Kepler-62f (top) and Kepler-186f (bottom). The power peaks are highlighted with red circles. Dashed lines are drawn where frequencies are predicted using the Lagrange-Laplace (L-L) approach (section \textsection \ref{ss:ll-th}) and follow the same color code as Figure \ref{f:5p-mad}. The L-L predictions show high consistency with the {\emph N}-body results overall. For Kepler-186f, the peak at $-223\arcsec yr^{-1}$ is an alias of the $-1520\arcsec yr^{-1}$ mode. We do not detect significant real peaks in the positive frequency regime, which agrees with the L-L analysis. }
\label{f:num-res}
\end{figure}

As explained in section \textsection \ref{ss:res-zones}, obliquity resonance occurs where the precession rate of the spin axis coincides with a modal frequency of the orbital inclination. Therefore, the properties of the orbital inclination oscillation modes can be used to predict the behavior of obliquity variations. Empirically, the modes can be obtained from a Fourier transform (FT) on the time series of orbital inclination modulated by the ascending node, $i(t) e^{\sqrt{-1}\Omega(t)}$. Using the {\emph N}-body results described in section \textsection \ref{s:nr}, we calculate the FTs for Kepler-62f and Kepler-186f in the near-coplanar cases and present them in Figure \ref{f:num-res} in black. The locations of peaks correspond to the modal frequency values, $f_k$, which determine the centers of the obliquity resonances via Eqn.(\ref{e:coseps}). The peak amplitude reflects the power associated with each mode, $i_k$, which in turn determines the width of the corresponding obliquity resonance through Eqn.(\ref{e:wid}). Figure \ref{f:num-res} focuses on the negative frequency regime because in neither scenario do we detect notable real structure (i.e. not attributable to aliasing) in the positive frequency regime. This explains our numerical finding that retrograde obliquities are stable for these configurations.  

For Kepler-62f, we identify two dominant nonzero frequency modes (amplitude $> 0\farcs01$), with frequencies $f_k = -10\farcs0 ~{\rm yr}^{-1}$ and $-41\farcs8~{\rm yr}^{-1}$, and FT peak amplitudes $i_k = 0\fdg09$ and $i_k = 0\fdg014$ for the coplanar configuration. 
Following Eqn.(\ref{e:coseps}), we can predict the locations of the resonant regions of $\cos(\epsilon)$ for given values of the precession coefficient $\alpha_p$. In particular, for an Earth-like $\alpha_p = 32\farcs2~{\rm yr}^{-1}$, we would expect Kepler-62f to exhibit a resonant region at $\epsilon = 71\fdg8$ induced by the $-10\farcs0~{\rm yr}^{-1}$ mode ($f_k = -41\farcs8~{\rm yr}^{-1}$ yields an unphysical $\epsilon_{\rm res}$ for this $\alpha_p$). Substituting the amplitude of the mode in Eqn.(\ref{e:wid}), the half-width of the resonance is expected to be $\sim 2\fdg6$. The half-amplitude measured from the numerical simulation is $\sim2\fdg3$ 

With a larger precession coefficient of $\alpha_p = 45\farcs5~{\rm yr}^{-1}$ ($P_{\rm rot} = 17$ hr), corresponding to the faster rotator as presented in the middle and lower left panels in Figure \ref{f:eps-t} for Kepler-62f, both modes yield physical resonant regions, at $\epsilon = 77\fdg3$ and $\epsilon = 23\fdg0$. In the coplanar case, the resonance half-widths are $2\fdg2$ and $2\fdg6$ respectively. The numerical results for the faster rotator show variability half-amplitudes at $\epsilon \sim 77^\circ$ and $\sim 23^\circ$ to be $1\fdg8$ and $2\fdg9$ respectively. Overall, these analytical results agree well with the numerical results shown in the top and middle left panels of Figure \ref{f:eps-t}. In general, the resonance widths given in Equation (\ref{e:wid}) are an upper limit for the variability amplitude because the extent to which the variability occupies the full width of the resonance depends on the exact initial obliquity, $\epsilon_0$, as well as the initial spin axis longitude, $\psi_0$. Underestimates by the analytical resonance-width equation using the sharp FT peak values as inputs could happen when the numerical FT does not resolve the modal frequencies, which underestimates the modal amplitudes.   

Increasing the orbital mutual inclinations tends to boost the amplitude $i_k$ associated with each inclination mode, which leads to enlarged resonant widths (\textsection \ref{ss:res-wid}). This phenomenon is exemplified in the {\emph N}-body results corresponding to the bottom panels in Figure \ref{f:eps-t}, where the orbital inclination of Kepler-62f reaches $\sim 1\fdg4$. While the modal frequencies remain almost unchanged from that of the near-coplanar case ($-10\farcs0~{\rm yr}^{-1}$ and $-41\farcs8~{\rm yr}^{-1}$), the modal amplitudes increase to $0\fdg32$ and $0\fdg17$, respectively (not shown in Figure \ref{f:num-res}). Thus, according to Eqn.(\ref{e:wid}), the corresponding half-widths of the resonances for the faster rotator increase to $4\fdg0$ and $8\fdg2$. Again, the location of the resonant regions agrees very well with the numerical results shown in section \textsection \ref{s:nr}. The numerical half-amplitudes are $2\fdg9$ and $10\fdg4$. Importantly, the expression for the half-width is no longer a good approximation when the variability is large (i.e. when $\epsilon$ deviate from $\epsilon_{\rm res}$ in the resonant region).

Since every dominant mode can in theory cause low-obliquity instability for an appropriate $\alpha_p$ value (see section \textsection \ref{ss:res-zones} and \textsection \ref{ss:ll-th}), we should expect Kepler-62f to have two rotation periods that lead to destabilized low-obliquity zones. One of the modes ($f_k=-41\farcs8 ~{\rm yr}^{-1}$) is responsible for the lower-obliquity instability at $P_{\rm rot} \sim 15-20$ hr shown in Figure \ref{f:eps-t}. The other mode ($f_k=-10\farcs0~{\rm yr}^{-1}$) results in a similar instability region at $P_{\rm rot} \sim 70$ hr.

Via the same method, for Kepler-186 we can only identify one dominant mode (with amplitude $>0\fdg01$) at $-11\farcs7 $yr, where the amplitude is $0\fdg39$. Then, for an Earth-like $\alpha_p = 137\farcs4 ~{\rm yr}^{-1}$, the resonance is located at $\epsilon = 85\fdg1$ and the half-width of the resonance is $2\fdg76$, according to the analytical approximation in Eqn.(\ref{e:wid}). From the numerical results, the half-amplitude is $1\fdg7$. For the slower rotator with $\alpha = 12\farcs7 ~{\rm yr}^{-1}$, the resonant region is at $\epsilon = 23\fdg0$ and its expected half-width is $11\fdg7$, while in the numerical tests the obliquity half-amplitude is $8\fdg8$. For the case with higher mutual inclinations, where the inclination of Kepler-186f reaches $\sim 1\fdg5$, the modal amplitude is $0\fdg72$, corresponding to a resonance half-width of $15\fdg2$, while the numerical half-amplitude is $9^\circ$. Similar to Kepler-62f, the analytical results of the resonant location are consistent with the numerical values (shown in Figure \ref{f:eps-t}), though the analytical approximation somewhat overestimates the amplitude of the obliquity variations for reasons outlined earlier. Another small peak occurs at $\sim -210\arcsec~{\rm yr}^{-1}$ and is actually an alias of a more distant mode at $-1520\arcsec~{\rm yr}^{-1}$ (see \textsection \ref{ss:ll-th} and Table \ref{t:table-ll}).\footnote{The Nyquist frequency for our {\emph N}-body sampling rate (1/1000 yr) is $-648\arcsec{\rm yr}^{-1}$. Therefore, frequencies outside of this window can manifest as aliases folded on this value. }

\bigskip
\subsection{Dependence on Planetary Rotation Rates and Observational Uncertainties}\label{ss:ll-th}

In addition to Fourier transforming the {\emph N}-body results, one can alternatively estimate the relevant fundamental forcing frequencies of the planetary systems using a completely analytical approach following the Laplace-Lagrange (L-L) secular theory \citep[e.g.][]{Murray99}, which allows us to estimate the characteristic oscillation frequencies $f_k$ and their approximate amplitudes $i_k$ in planet $p$'s orbital plane induced by the other planets in the system. This is done by solving for the eigenvalues and eigenvectors of a matrix constructed from the masses and orbital semi-major axes present in the planetary system. The L-L approximation is accurate for systems with no mean motion resonances and containing nearly circular and coplanar orbits, conditions that are satisfied here. Such an approach is powerful because it allows us to rapidly characterize vast swaths of parameter space and visualize the results.

\begin{figure}[htb]
\begin{minipage}{0.5\textwidth}
\centering
\includegraphics[width=3.7in]{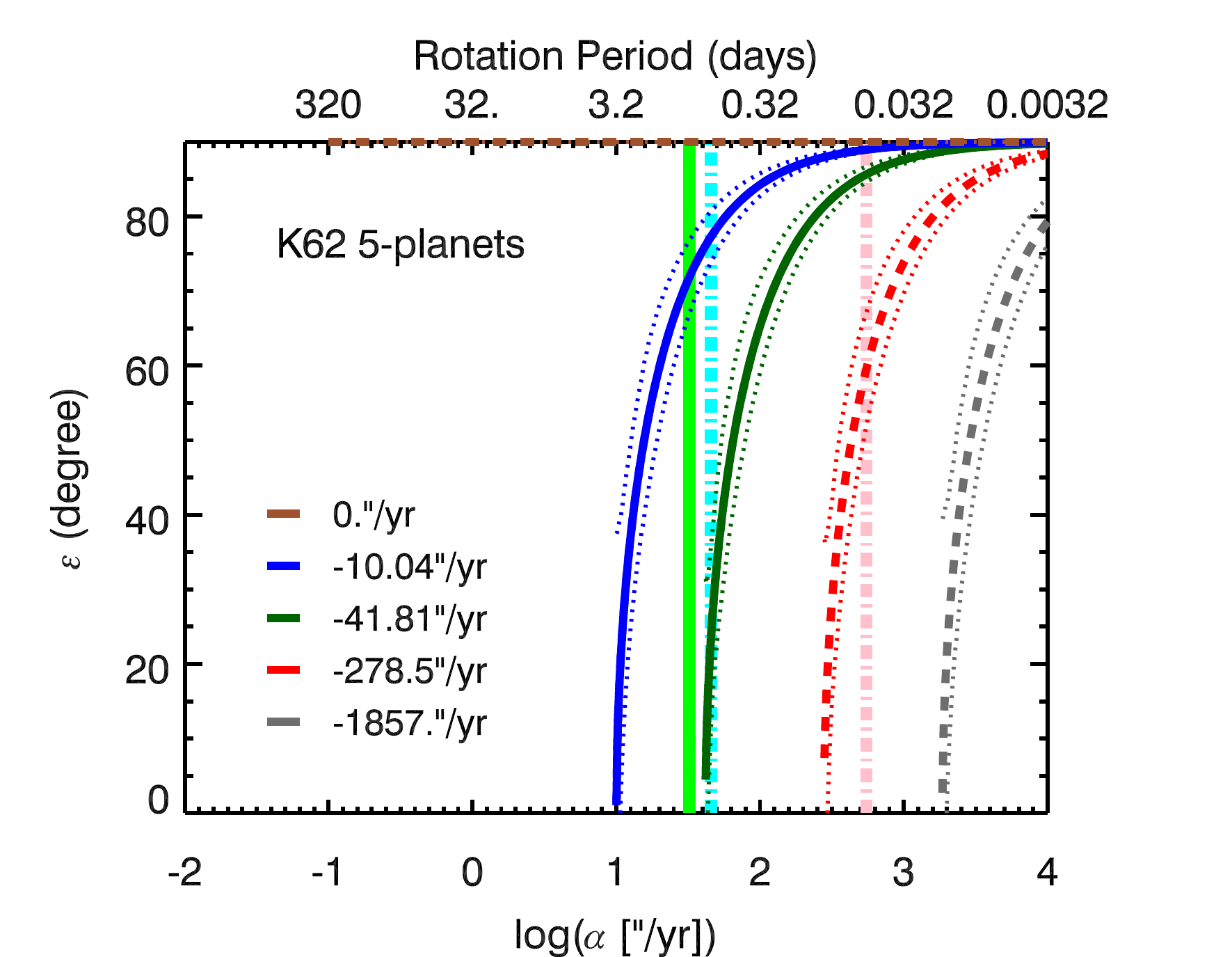}
\end{minipage}%
\begin{minipage}{0.5\textwidth}
\centering
\includegraphics[width=3.7in]{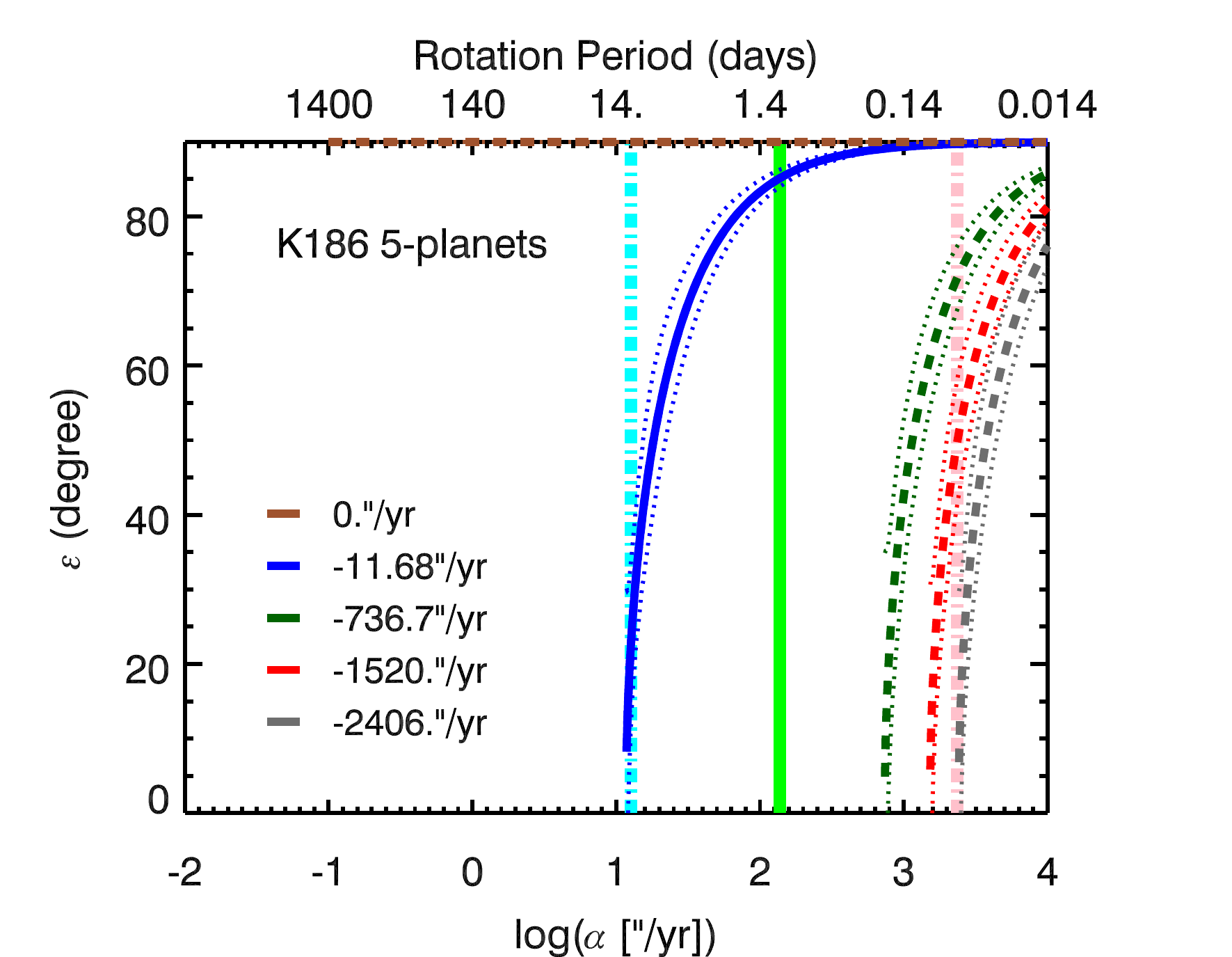}
\end{minipage}
\caption{Resonant obliquities as a function of precession coefficient ($\alpha$) for Kepler-62f (top panel) and Kepler-186f (bottom panel). Each curve corresponds to a modal frequency on orbital inclination excited by the planetary bodies in the system (as labeled). Solid curves depict dominant modes (i.e. $i_k > 0\fdg01$) while minor modes are shown with dashed curves. The MAD ranges for each frequency are plotted as dotted curves around the thick solid curves, based on Monte Carlo simulations over 1000 system realizations varying the stellar and planetary masses within their measurement uncertainties. The green solid vertical line represents the $\alpha_p$ values applicable if the planets were Earth analogs with $P_{\rm rot} = 1$ day. The cyan dot-dotted line denotes the planets with hypothetical rotation rates illustrated in Figure \ref{f:eps-t}. The pink dot-dotted line corresponds to breakup velocity. }
\label{f:5p-mad}
\end{figure}

We include the $f_k$ values computed using the L-L approach in Figure \ref{f:num-res}, shown by vertical dashed lines following a color code for direct comparison to Figure \ref{f:5p-mad}. The modal properties are also summarized in Table \ref{t:table-ll}. Notice that, consistent with Section \textsection \ref{ss:num-freq}, no positive frequencies exist, thereby explaining why retrograde obliquities have all been stable. According to the L-L approach, the most dominant modes for Kepler-62f are $f_k = -41\farcs8 ~{\rm yr}^{-1}$ and $f_k = -10\farcs0~{\rm yr}^{-1}$, with normalized eigenvector elements of $0.56$ and $0.29$ respectively.\footnote{The magnitude of the eigenvector element corresponding to a given mode is an indicator for its relative dominance.}  Solving for the amplitudes of the modes applying the same initial condition as those for the numerical simulations, we found the amplitudes of the dominant modes to be $i_k \sim 0\fdg015$ and $i_k \sim 0\fdg14$ respectively, which agree rather well with the results from the numerical Fourier transform. As discussed in section \textsection \ref{ss:num-freq}, the numerical peaks are typically underestimated due to the finite resolution of the FT. For Kepler-186f, $f_k = -11\farcs6~{\rm yr}^{-1}$ is the most dominant mode, with normalized eigenvector elements of $0.80$. Assuming the same initial condition as those for the numerical simulation, the modal amplitude is $i_k \sim 0\fdg39$, which is precisely what is observed in the Fourier transform. Therefore, there is good agreement on the behavior of the obliquity variations and the locations of the instability regions between the numerical results and the analytical ones. L-L uncovers additional nonzero modes that do not directly manifest in the FTs in section \textsection \ref{ss:num-freq} because they are outside the Nyquist frequency of the {\emph N}-body data sampling (except in the form of aliases). In any case, these missed modes are exceedingly minor and hence contribute negligibly to variability. 

In the high-inclination examples depicted in Figure \ref{f:eps-t}, we observed small-scale variabilities between the resonant zones predicted by L-L. These small-scale variabilities do not correspond to any primary resonances. It is important to remember that L-L assumes that the system has low orbital inclination and eccentricity values. As configurations deviate from the small-inclination and small-eccentricity regime, higher-order effects can become more prominent, leading to higher-order resonant regions. For instance, Figure \ref{f:num-res} illustrates hints of FT peaks missed by L-L even in the near-coplanar regime. These higher-order peaks grow as orbital mutual inclinations increase, and they are responsible for the small variabilities away from the L-L resonances.

\setlength{\tabcolsep}{4pt}
\begin{table}[tbh]
\begin{center}
\caption{Modal properties based on L-L method}
\label{t:table-ll}
\begin{tabular}{lcccc}
\hline
\multicolumn{5}{c}{\bf{Kepler-62f }}\\
\hline
\hline
$f_k$ ($\arcsec/yr$) & $\hat{I}_{k,5}$ & $I_{k,5; cop}$ ($^\circ$) & $I_{k,5; h}$ ($^\circ$) & $\epsilon_{\rm res}$ ($^\circ$) \\
\hline               
0 		& 0.45 			 &  0.11 				& 0.56 & 90   \\                         
-10.0 	& -0.29 				& -0.14   				& -0.46  & 72  \\                                                                                
-41.8 	& 0.56 				& 0.015				& 0.22  & \nodata \\                          
-279 	& -$3.5\times10^{-4}$ 	& $-1.9\times10^{-4}$ 	& $-8.0\times10^{-4}$ & \nodata \\                                
-1857 	& $2.2\times10^{-6}$ 	& $1.54\times10^{-7}$  	& $3.8\times10^{-6}$ & \nodata \\ 
\hline
\multicolumn{5}{c}{\bf{Kepler-186f }}\\
\hline
\hline
$f_k$ ($\arcsec/yr$) & $\hat{I}_{k,5}$ & $I_{k,5; cop}$ ($^\circ$) & $I_{k,5; h}$ ($^\circ$) & $\epsilon_{\rm res}$ ($^\circ$) \\
\hline               
0 		& 0.45 				&  0.35 				& 0.73 & 90   \\                         
-11.6 	& 0.80  				& 0.39	   			& 0.72 & 85.2  \\                                                                                
-737		& 0.0027 				& $1.1\times10^{-3}$ 				& $7.7\times10^{-3}$  & \nodata \\                          
-1520 	& $2.1\times10^{-4}$ 	& $1.1\times10^{-4}$ 	& $1.9\times10^{-4}$ & \nodata \\                                
-2407	& $2.9\times10^{-6}$ 	& $1.0\times10^{-6}$  	& $2.9\times10^{-6}$ & \nodata \\ 
\hline
\hline
\end{tabular}
\vspace{0.25cm}
\end{center}

Note. Column headings:
$\hat{I}_{k,5}$ -- normalized eigenvector elements; $I_{k,5; {\rm cop}}$ -- components of the inclination eigenvector, the absolute values of which characterize the size of $i_k$ values for the near-coplanar configuration; $I_{k,5; h}$ -- same as $I_{k,5; {\rm cop}}$, but for the higher mutual inclination ($< 3^\circ$) configuration;
$\epsilon_{\rm res}$ -- center of obliquity resonance region corresponding to Earth-like $\alpha_p$.

\end{table}  

As shown in Eqn.(\ref{e:alpha}), the precession coefficient $\alpha_p$ is highly sensitive to the planet's spin rate, as it determines the oblateness of the planet. To comprehensively investigate the dependence of obliquity variation as a function of the planetary spin rate, we calculate the resonant obliquity values as a function of the precession coefficient $\alpha_p$ using Eqn.(\ref{e:coseps}). The result is shown in Figure \ref{f:5p-mad} for Kepler-62 and Kepler-186. We label the corresponding planetary rotation period in the top x-axis of Figure \ref{f:5p-mad}, assuming $\alpha_p \propto \omega$ as discussed in Section \ref{ss:ham}. In general, ignoring the oblateness of the host star, a system of $N$ planets will induce $N-1$ coupled nonzero frequency inclination modes. The colored solid curves represent the resonant obliquities corresponding to the five different inclination modes (including $f=0$). For a given value of $\alpha_p$, only a subset of the modes will correspond to physical values of $\epsilon_{\rm res}$. The valid resonant obliquities appear in Figure \ref{f:5p-mad} as intersections between an $\alpha_p$ vertical and the curves. Chaotic zones, where the obliquity evolution becomes unpredictable, occur when two resonant zones overlap with each other, which becomes more likely where the resonant widths are large and curves are dense.

The values of $f_k$ depend on the stellar mass $M_\star$, as well as the masses and semi-major axes of each planet $j$ in the system ($m_j, a_j$). While the orbital period is measured with high precision, the masses are often underconstrained. To investigate the sensitivity of the resonance locations to measurement errors, we conduct Monte Carlo experiments over $1000$ system realizations to obtain a measure of the error in the determinations of the median resonant obliquity. Each realization is assigned a stellar mass and a radius for each planet. Both are drawn from Gaussian distributions defined by their respective medians and errors quoted in Section \textsection \ref{s:sys-pars}. To generate the planet mass, each planet radius draw is fed into {\tt{Forecaster}} \citep{ChenKip17}, which outputs a draw from the mass posterior, taken to be $m_p$ for that realization. The exceptions are Kepler-62f ($3.3^{+2.3}_{-0.6} M_\earth$) and Kepler-186f ($1.7^{+1.1}_{-0.3} M_\earth$), for which the mass posteriors are calculated from the Terran power law marginalized over the hyperparameter distributions. Essentially, such a procedure marginalizes over the full planetary mass distributions accounting for uncertainties in radius. 

In Figure \ref{f:5p-mad}, the median absolute deviation (MAD) of the resonant obliquities are overplotted in dashed lines. The MADs of the modal frequencies only vary within $\sim 20-30\%$ of their median value, and the modes with frequencies close to $-41\farcs8~{\rm yr}^{-1}$ and $-10\farcs0~{\rm yr}^{-1}$ ($-11\farcs6~{\rm yr}^{-1}$) remain dominant for Kepler-62f (Kepler-186f) in the Monte Carlo simulations. It appears that the indeterminacy in these curves are relatively small in the log plots. 

In Figure \ref{f:5p-mad}, we use green solid lines to indicate the precession frequency corresponding to Earth analogs (i.e. $P_{\rm{rot}} = 1$ day). Both Kepler-62f and Kepler-186f avoid instability zones at Earth-like obliquity values (rather narrowly for Kepler-62f). However, one can expect higher obliquities to undergo mild variations. Any moderate mutual orbital inclination could also enlarge the widths of each resonant zone. These predictions are consistent with the numerical results in Section \ref{s:nr}. We also overplot with cyan dot-dotted lines $\alpha_p$ values corresponding to the bottom panels of Figure \ref{f:eps-t}, and show that they coincide with regions where lower obliquities undergo resonances. This picture indicates that the resonant regions in the low-obliquity region is highly fine-tuned in rotation period. 

One rotation rate of interest is that corresponding to the planet's breakup velocity, which is 0.0586 days for Earth analogs (shown as pink dot-dotted lines in Figure \ref{f:5p-mad}). Rotation periods shortward of this value are unphysical. If rotating close to breakup, both Kepler-62f and Kepler-186f intersect with multiple resonant curves at small obliquities. Nevertheless, since the resonant amplitudes are low for these modes (see Table \ref{t:table-ll}), the associated variability would be quite limited. Of course, given the probable ages of the systems and tidal interactions over time, it is unlikely that the planets are such extreme rotators today 
anyway. Higher-precession coefficients than the one corresponding to the breakup velocity are allowed for planets with moons (see section \textsection \ref{ss:moon}). 

Another regime of relevance in the long term is that of synchronized rotation with the orbital periods of 267 days and 130 days for Kepler-62f and Kepler-186f, respectively (see also Section \ref{ss:tides}). The synchronized rotation periods do not correspond to physical resonant obliquity values. As pointed out by \citet{Bolmont14, Bolmont15} and \citet{Shields16}, planetary obliquities eventually settle to 0, though the rate at which tidal synchronization occurs is dependent on many factors, such as the planetary tidal $Q$-values \citep{Heller11}. Kepler-62f and Kepler-186f have likely not yet reached this state \citep[][see also Section \ref{ss:tides}]{Bolmont14, Bolmont15, Shields16} but are inevitably marching toward it. From Figure \ref{f:5p-mad} it is visually apparent that, en route to the synchronized states, the spin rate and therefore the precession frequency of the planet decline and allow the system to sweep across the resonant regions. This allows obliquity variations in a wide range of obliquity regions similar to the future evolution of the Earth as discussed by \citet{deSurgy97}.

\section{Discussion}\label{s:dis}

Accurate determination of many properties of the exoplanets remains out of reach with current techniques. As will be discussed below, a planet's obliquity evolution can be affected by many factors that are not well constrained. This includes the presence of additional planets and satellites, as well as the rotation periods and the oblateness of the planets. Therefore, the parameter space governing a planet's obliquity evolution is large, and to accurately pinpoint the location occupied by a given exoplanet is challenging.

What is possible is to map out representative regions of this parameter space and infer the general behaviors. Such characterization could provide guidance over the range of considerations toward assessing the planet's obliquity evolution. In this section we investigate the effect of the existence of additional planets, including giant Jupiter/Saturn analogs as well as internal rocky planets. We also consider the influence of a satellite, which plays a critical role in stabilizing the Earth's own obliquity \citep{Laskar93a,deSurgy97}. Finally, we discuss the planets' long-term obliquity dynamics as it gradually synchronizes its rotation with its orbital period as a result of tidal interactions.     

\bigskip
\subsection{Extra Planets}\label{ss:more-p}

Thus far, our analysis has been anchored on the assumption that all the relevant planets that exist in the systems have been detected. In reality, the sensitivity of the transiting technique falls with increasing planet distance or inclination. The presence of additional planets, especially at high orbital inclinations, can influence the spin dynamics in a given system dramatically through the introduction of additional, potentially strong modes. In this subsection we map the resonant zones for system configurations involving external giant planets (\textsection \ref{sss:ext-p}), as well as an internal nontransiting planet (\textsection \ref{sss:int-p}). 

\subsubsection{External Giant Planets}\label{sss:ext-p}

The current estimate on the occurrence rate of Jupiter-massed planets orbiting at Jupiter-like distances around M dwarfs is $\sim6\%$ \citep{Clanton16, Meyer17}. Therefore, while atypical, external giant planets are not wholly uncommon in these systems. Jointly motivated by the architecture of our own solar system, we characterize the influence of hypothetical external giant planets on the obliquity evolution of Kepler-62f and Kepler-186f using the analytic techniques outlined in previous sections, and we confirm them with numerical integration. 

\begin{figure}[htb]
\begin{minipage}{0.5\textwidth}
\centering
\includegraphics[width=3.7in]{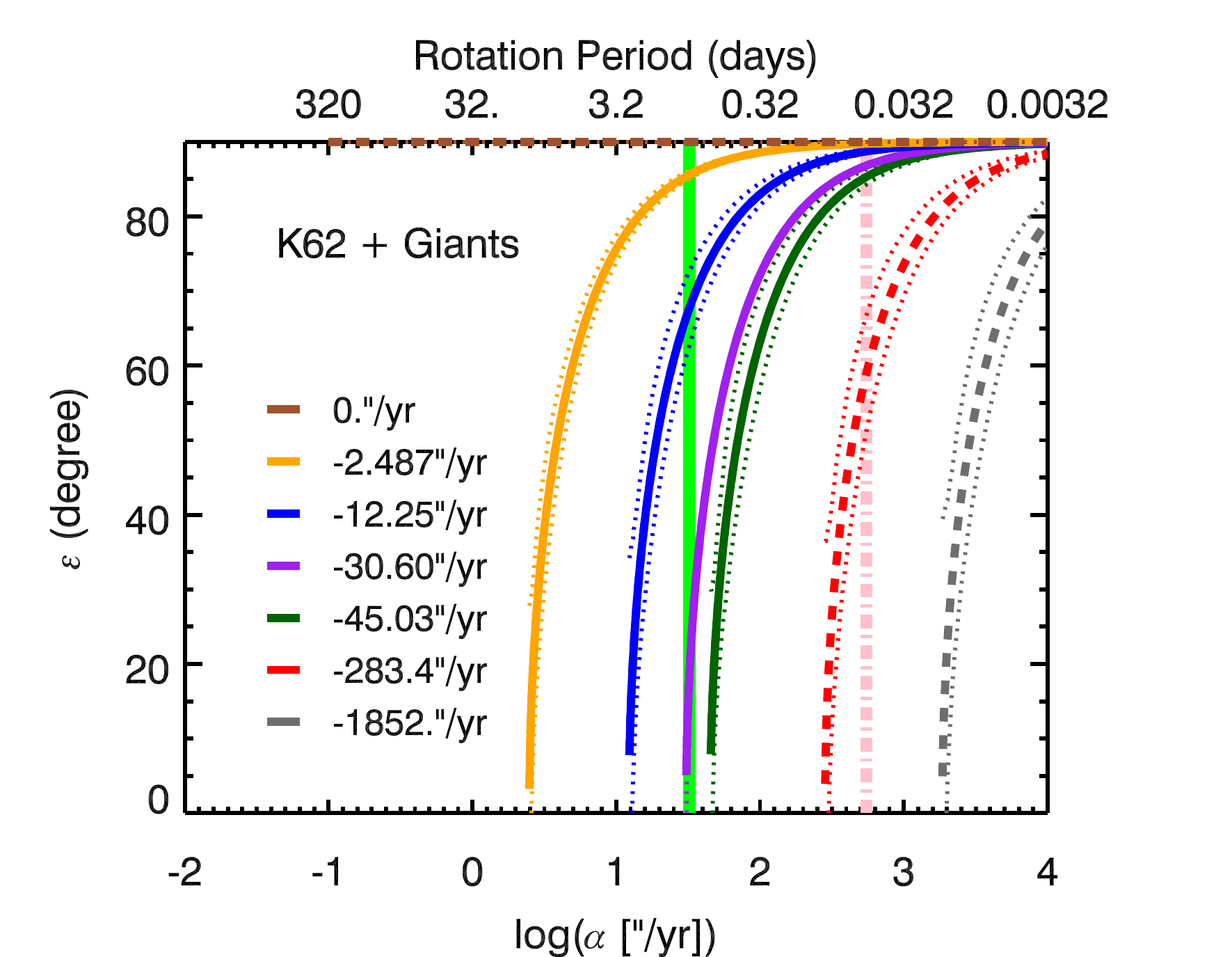}
\end{minipage}%
\begin{minipage}{0.5\textwidth}
\centering
\includegraphics[width=3.7in]{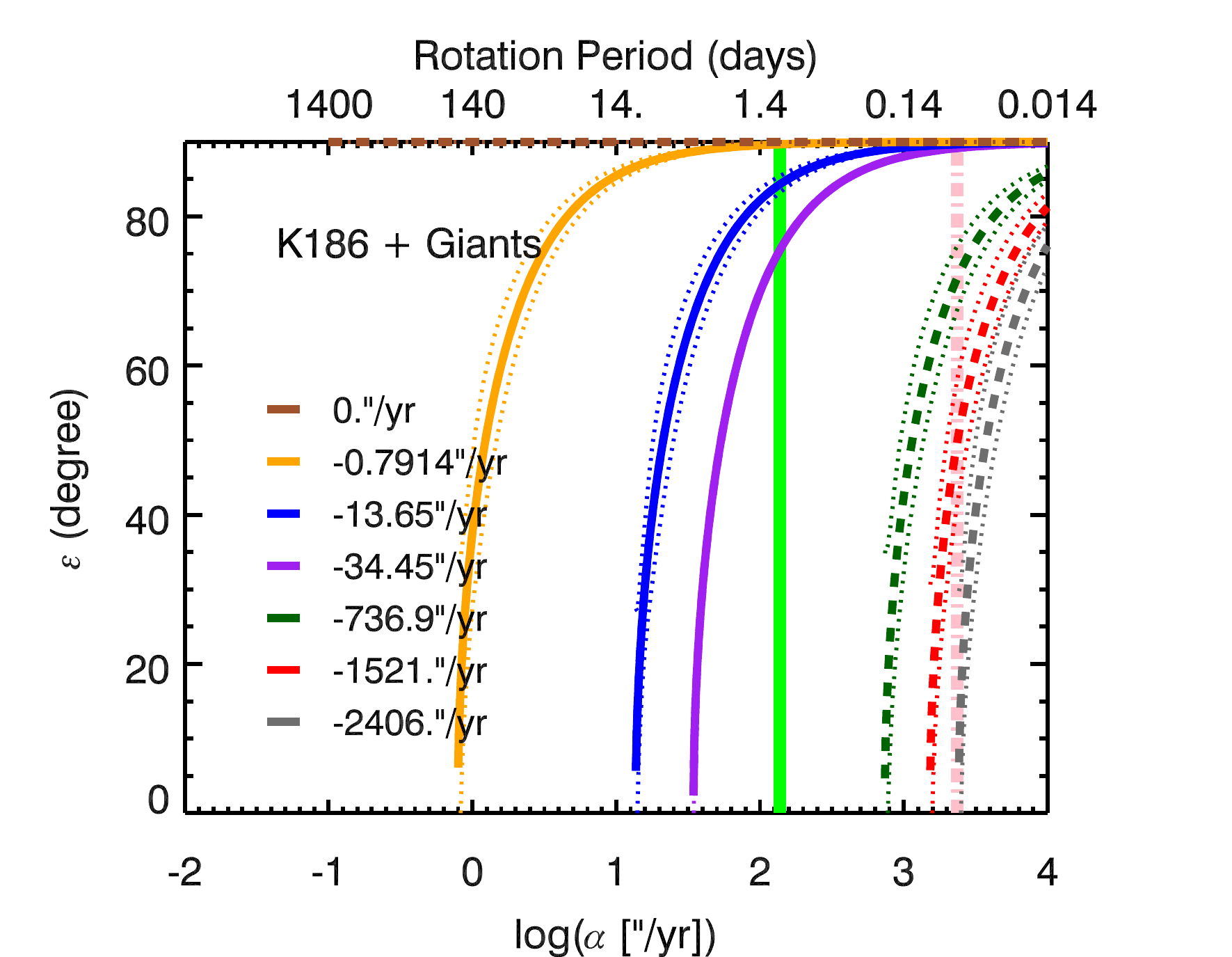}
\end{minipage}
\caption{Same as Figure \ref{f:5p-mad}, but now with distant giant planets in the form of Jupiter and Saturn analogs. The additional frequencies induced tend to be low and impact slow rotators more severely, but they also modify the locations of the other resonances. In this scenario, Kepler-62f with an Earth-like precession coefficient can expect to experience chaotic obliquity evolution even in the low-obliquity regime. }
\label{f:7p-mad}
\end{figure} 

As an example, we place a Jupiter analog and a Saturn analog in each system and calculate the resonant obliquities induced by each using the L-L theory. In this context, `analog' signifies a planet with identical mass, orbital inclination, and a similar semi-major axis, in the present day. For Jupiter and Saturn, the orbital distances are $\sim 5.2$ and $9.6$ au and their inclinations are small ($1\fdg304$ and $2\fdg485$). Including only a Jupiter-analogue, a new dominant inclination variation mode is introduced at $f_k \sim -2\farcs5~{\rm yr}^{-1}$ for Kepler-62f, and at $f_k \sim -0\arcsec8~{\rm yr}^{-1}$ for Kepler-186f. The corresponding normalized eigenvector elements for this mode are $\sim 0.3$ for Kepler-62f and $\sim 0.4$ for Kepler-186f. These results are qualitatively similar for a range of orbital distances of Jupiter between 3 and 7 au. Including both the Jupiter and Saturn analogs, there is another mode introduced at $f_k \sim 30 \arcsec ~{\rm yr}^{-1}$ for Kepler 62f and $f_k \sim 34 ~{\rm yr}^{-1}$ for Kepler 186f. The modes attributed to the Saturn-analogue are much weaker than those introduced by Jupiter, with normalized eigenvector elements of $\sim 0.04$ for Kepler-62f and $\sim 0.03$ for Kepler-186f.

The L-L resonant curves in the $\alpha$(or $P_{\rm rot}$)-$\epsilon$ plane are shown in Figure \ref{f:7p-mad} for systems including both the Jupiter and the Saturn analogs, where the orbital distances are fixed to those of Solar System's Jupiter and Saturn. The solid lines represent the default orbital parameters as presented in Table \ref{t:sys-pars}, and the dashed lines represent the MAD values based on the Monte Carlo simulations taking into account the observational uncertainties of the planetary masses. An Earth analog in such systems would encounter more obliquity instability zones occurring at lower obliquities. For Kepler-62f, the resonant regions caused by the dominant mode $f_k \sim -45\farcs0~{\rm yr}^{-1}$ and the less dominant mode introduced by Saturn $f_k \sim 30\farcs6~{\rm yr}^{-1}$ are only slightly separated in the low-obliquity regime. Thus, one might expect the obliquity evolution of such a planet to be chaotic owing to higher probability of overlap between the resonances. The other dominant modes for both Kepler-62f and Kepler-186f correspond to low frequencies, which implies a greater chance for instability and chaos at higher obliquity angles for Earth analogs, as well as more severe impact on lower-obliquity regions for slowly rotating planets.  

\begin{figure}[htb]
\centering
\vspace{-1cm}
\hspace{-0.5cm}
\includegraphics[width=3.6in]{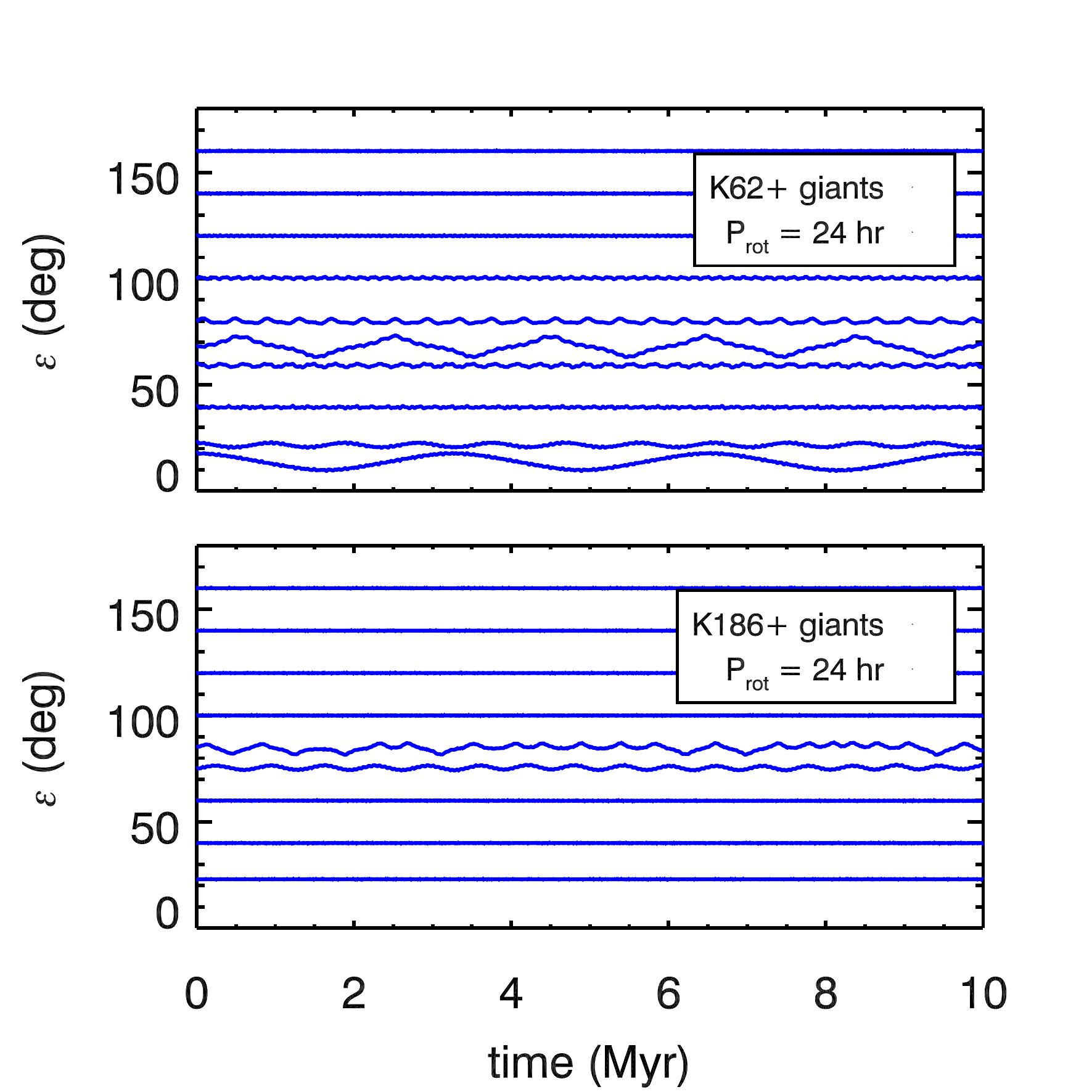}
\caption{Obliquity variation of Kepler-62f (top panel) and Kepler-186f (bottom panel) with $P_{\rm rot} = 24$ hr in the presence of outer Jupiter- and Saturn-like planets. Variations can be large in the lower-obliquity range for Kepler-62f, consistent with the analytical estimates. For Kepler-186f with $\epsilon \sim 85^\circ$ in this hypothetical system, the obliquity evolution is chaotic owing to the overlap of the two resonant zones associated with the $-0\farcs8~{\rm yr}^{-1}$ and $-13\farcs6~{\rm yr}^{-1}$ frequencies. In all cases, retrograde obliquities are stable. }
\label{f:obSJ}
\end{figure} 

For comparison, we numerically compute the obliquity evolution for Earth analogs (i.e. $P_{\rm rot} = 24$ hr) in systems including both Jupiter and Saturn analogs, and we present them in Figure \ref{f:obSJ}. The inner planets start nearly coplanar to each other and are stirred into more inclined configurations by the giant planets, which start with their present solar system orbital inclinations ($1\fdg304$ and $2\fdg485$). Consistent with the analytical expectations, the lower-obliquity regions of Kepler-62f allow larger variabilities owing to two closely spaced inclination oscillation modes. The higher-obliquity region also exhibits larger amplitude variabilities than the case without the outer giant planets (as shown in Figure \ref{f:eps-t}). This is because the giant planets induce larger mutual inclinations between all the planets. For Kepler-186f, the variabilities at low obliquities are still low, due to the lack of additional resonant regions in that region. Similar to Kepler-62f, high-obliquity variation is enlarged by the presence of the giant planets. Interestingly, an example of chaotic obliquity evolution is found around $85^\circ$, which is likely caused by some overlap of the resonant zones associated with the $-0\farcs8~{\rm yr}^{-1}$ and $-13\farcs6~{\rm yr}^{-1}$ frequencies. In all cases, retrograde obliquities are stable.  \citet{Quarlesinprep} also investigate the influence of outer giant planets on the spin axis variability of Kepler-62f. Our results are mostly consistent with their conclusions, except that \citet{Quarlesinprep} find obliquity variability in the retrograde regime for a specific realization of the system.

\bigskip

\subsubsection{Additional Planet Interior to Kepler-186f}\label{sss:int-p}

In addition to distant planets, it is also possible to have internal nontransiting planetary companions. In particular, the separation between Kepler-186e and Kepler-186f is large, and it is likely that an extra undetected planet exists between these two known planets, based on accretion disk simulations \citep{Bolmont14}. Using dynamical simulations, \citet{Bolmont14} characterized the mass of the extra planet. Considering a planetary mass ranging from $0.1M_{\oplus}$ to $1M_{Jup}$, \citet{Bolmont14} found that the planet cannot be more massive than the Earth, in order to keep the mutual inclinations between the rest of the planets low so as to allow them to transit. Here we adopt the orbital configuration of the extra planet assumed by \citet{Bolmont14} to study the obliquity evolution of Kepler-186f in the presence of this additional, more inclined body interior to its orbit, where $a_{\rm ex} = 0.233$ au, $e_{\rm ex} = 0.01$ and $i_{\rm ex} = 2^\circ$ to avoid transiting. We assumed the extra planet to be Earth-massed ($1M_{\oplus}$).

\begin{figure}[htb]
\begin{minipage}{0.5\textwidth}
\centering
\includegraphics[width=3.7in]{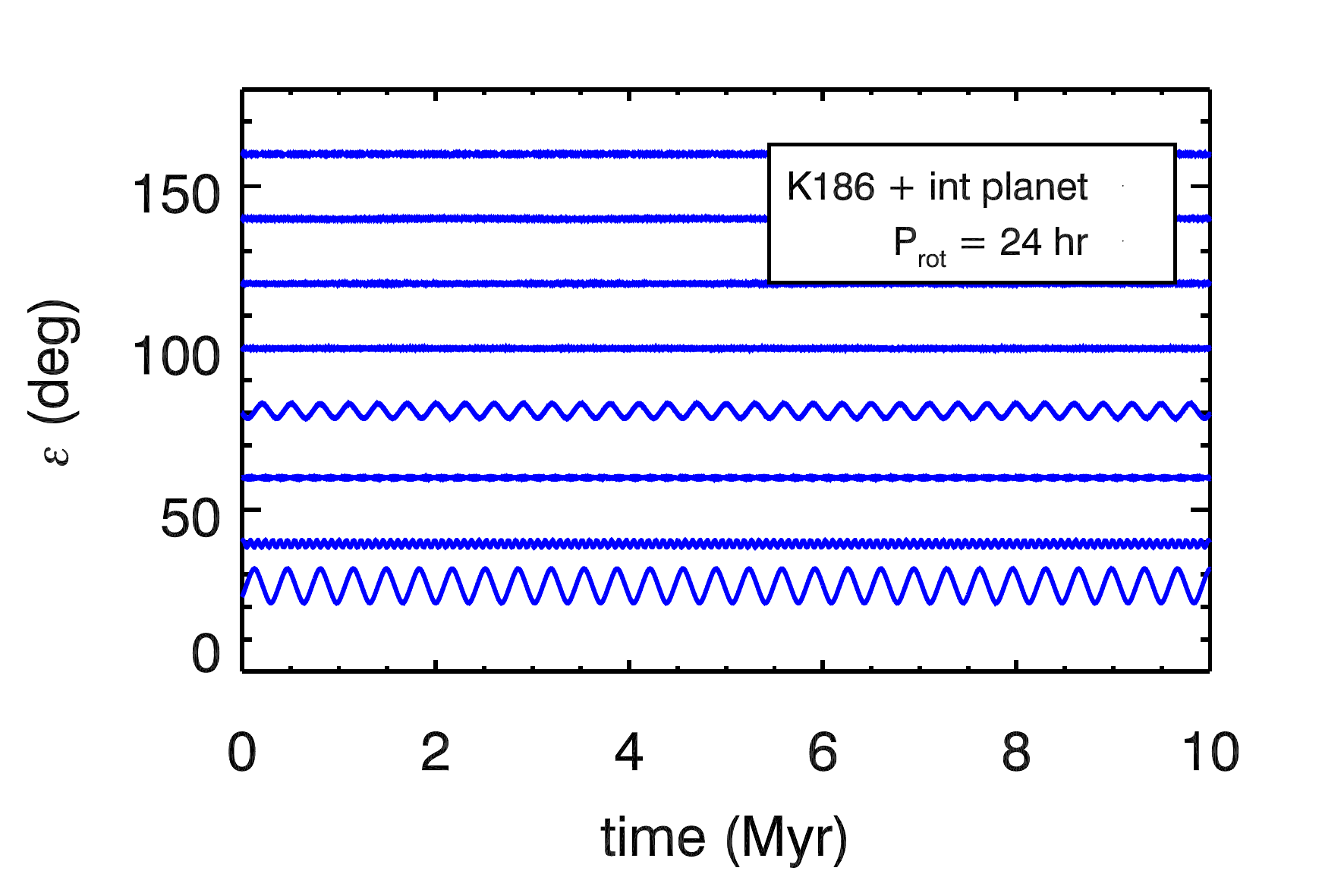}
\end{minipage}%
\caption{Obliquity evolution of Kepler-186f in the presence of an additional planet ($1 M_\earth$ between Kepler-186f and Kepler-186e. The low-obliquity region allows moderate variabilities, and the obliquity variation around $\sim 80^\circ$ is larger than the case without the additional planet.}
\label{f:eps186expl}
\end{figure}

To initialize the simulation, we adopt the nearly coplanar configuration of Kepler-186, as discussed in section \ref{s:nr}, and set the precession coefficient for planet f to be $\alpha = 137\farcs4~{\rm yr}^{-1}$, assuming that the planet is Earth-like with a rotation period of 24 hr. The obliquity evolution of Kepler-186f is shown in Figure \ref{f:eps186expl}. The low-obliquity region allows some variabilities, since the extra planet introduces a modal frequency at $f = 122\farcs5~{\rm yr}^{-1}$, which leads to a resonant region at $\epsilon \sim 26\fdg9$. 

Comparing with the obliquity variation without the extra planet (top right panel of Figure \ref{f:eps-t}), the variability in the high-obliquity region is also larger. This is because the higher inclination of the extra planet also excites the mutual inclination between the planets, which leads to a stronger perturbation to the planetary spin axis. The increase in the obliquity variation due to the extra planet is consistent with the discussions by \citet{Bolmont14}.

\bigskip
\subsection{Presence of a Satellite}\label{ss:moon}

The solar system is teeming with moons. However, little is known about moons elsewhere, in part due to the challenges associated with their low expected signal. Thorough searches in transiting exoplanet data have been conducted \citep[e.g.][]{Kipping13} but thus far have revealed only one possible exomoon candidate \citep{Teachey17}. \citet{Sasaki14} suggest that the tidal decay lifetimes of typical large moons around habitable planets of smaller stars tend to be shorter. However, for Kepler-62f and Kepler-186f, this timescale should exceed the best estimated system ages \citep{Shields16}. 

In any study of planetary obliquity evolution, it is important to consider the possible presence of moons, since moons cause additional torque to the planet's spin axes, hence increasing the precession coefficient. The impact could be large and is described by the modified version of Eqn.(\ref{e:alpha}) given below \citep[e.g.][]{deSurgy97}: 

\begin{align}
\label{e:alpha-moon}
\alpha_p = & \frac{3G}{2\omega}\biggl[\frac{M_\star}{(a_p\sqrt{1-e_p^2})^3}  \\
 + & \frac{m_M}{(a_M\sqrt{1-e_M^2})^3}\left(1-\frac{3}{2}\sin^2 i_M \right) \biggr]E_d. \nonumber
\end{align}

\noindent In the second term, $m_M$ is the mass of the moon. Similarly, the subscript $M$ on the orbital elements $a$, $e$, and $i$ indicates that these quantities pertain to the moon's orbit around the planet. Each additional moon contributes one such term, and their collective effect is additive. 

\begin{figure}[htb]
\hspace{-1cm}
\includegraphics[scale=0.55]{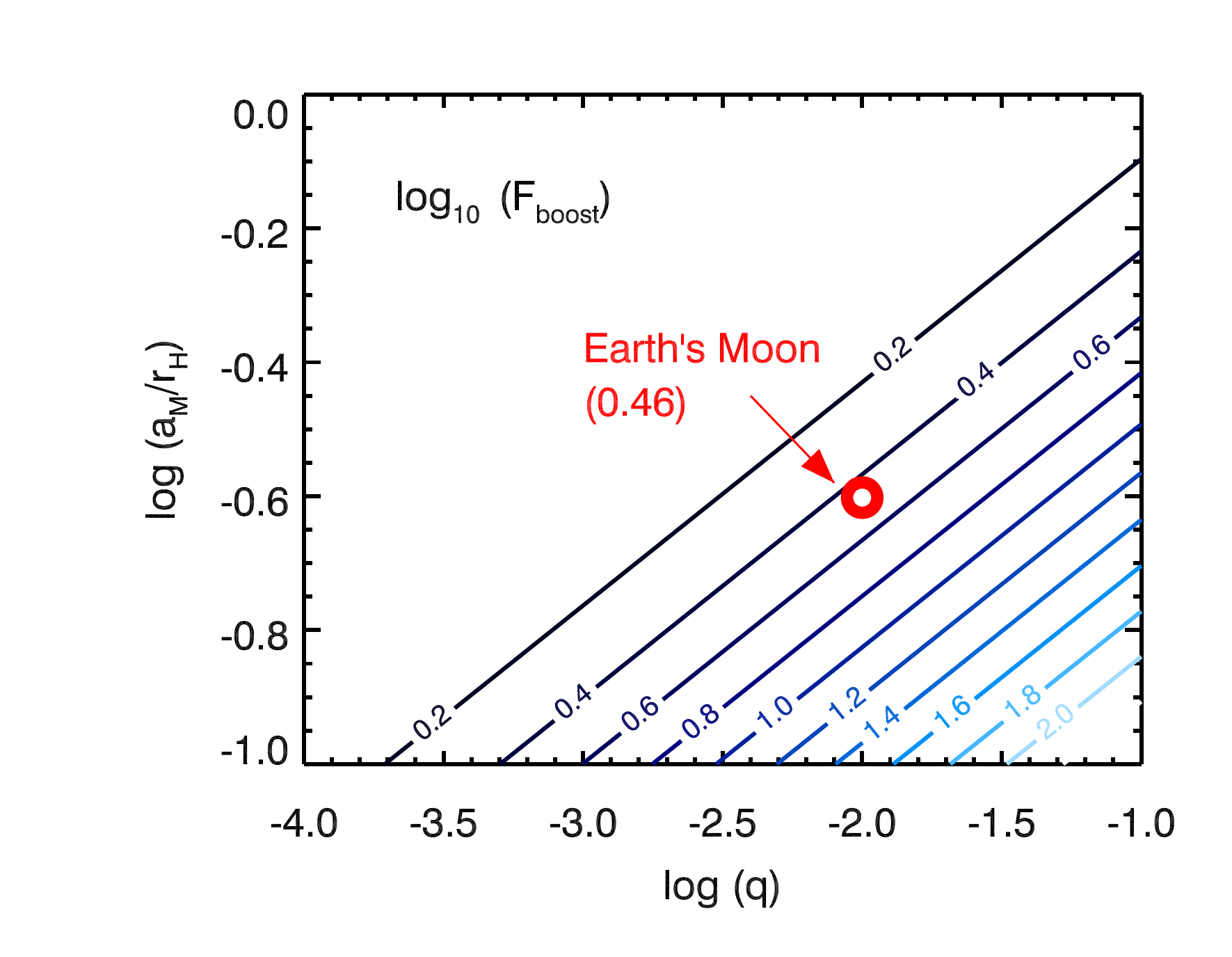}
\caption{The $\alpha_p$ enhancement factor, $F_{\rm boost}$, provided by a moon with mass $m_M = qm_p$ orbiting the planet on a circular, coplanar orbit at $a_M = fr_H$. Contours shown are $\log F_{\rm boost}$.  }
\vspace{+0.2cm}
\label{f:moon-alpha}
\end{figure}

For the Earth, the precession due to the Moon is about twice that of the Sun. The same is true for Kepler-62f and Kepler-186f, if we assume that each harbors a moon analogous to that of Earth, that is, a satellite that preserves the mass ratio and orbits at the same fraction of the planetary Hill radius ($r_{\rm H}$) as that in the Earth-Moon system. For a given planet-star pair, a moon for which $q = m_M/m_p$ and $f = a_M/r_H$ boosts the moonless version of $\alpha_p$ by the following factor: 

\begin{equation}
F_{\rm boost} (q,f) \equiv \frac{\alpha_M}{\alpha_\star} = 1 + 3 \frac{q}{f^3}, 
\label{e:moon-boost}
\end{equation}

\noindent assuming $e_M$ and $i_M \ll 1$. Here $\alpha_{\rm Moon}$ and $\alpha_\star$ refer to $\alpha_p$ with and without the moon, respectively. Figure \ref{f:moon-alpha} shows contours of $\log F_{\rm boost}$ in $\log f$ and $\log q$. The boost is enhanced at low $f$ and high $q$-values.      

The Earth-Moon arrangement turns out to be critical for the obliquity stability of the Earth. Without the Moon, the Earth sits in a large obliquity resonant zone spanning $0^\circ$ to $45^\circ$ \citep{Lissauer12,Li14}. Our Moon pushes the Earth away from this hazardous region \citep{Laskar93a}. 
Figure \ref{f:moon-alpha} shows that, in general, satellites with sufficiently large $q$ and/or small $a_M$ can push $\alpha_p$ beyond all the major frequencies present in the system, thereby stabilizing obliquity. This is helped by the fact that there exists upper bounds on the forcing frequencies from giant planets on HZ terrestrial planets due to orbital stability criteria \citep{Atobe04}. However, we point out that the same satellites could destabilize the obliquity for sufficiently slowly rotating planets, i.e. those associated with precession frequencies that are lower than some of the perturbing frequencies. Therefore, in general, changes in the lunar parameters could be fortuitous, catastrophic, or simply inconsequential, depending on the landscape of resonant obliquity regions (as illustrated in Figure \ref{f:5p-mad} and \ref{f:7p-mad}) and the $\alpha_p$ and $F_{\rm boost}$ involved. Moreover, a satellite could alter the rotation rate of the planet over time due to tidal interactions, as our Moon has done to Earth, which would further modify the oblateness of the planet, resulting in different $\alpha_p$ values (see \textsection \ref{ss:tides} as well as a detailed investigation in \citet{Atobe07}).

Assuming that Kepler-62f is an Earth analog with rotation period of $24$ hr, the increase to its $\alpha_p$ due to the existence of moons could destabilize its obliquity, if the new precession rate leads to resonant interactions with the inclination oscillation mode of $f_k = -41\farcs8~{\rm yr}^{-1}$. On the other hand, the precession coefficient of Kepler-186f with $P_{\rm rot} = 24$ hr is already larger than that of the frequency of the dominant inclination oscillation mode. Thus, the existence of a moon would not cause obliquity variations for Kepler-186f.





\bigskip
\subsection{Long-term Tidal Evolution}\label{ss:tides}
Tidal locking implies synchronization between a planet's rotation and orbital periods, as well as the direction of its spin and orbit vector. Both Kepler-62f and Kepler-186f are relatively far from their host stars, thereby likely to have weak tidal interactions with their host stars. However, whether or not they have already become tidally locked depends sensitively on the detailed properties of the planets, as well as the amount of time available to synchronize. Such incertitude has been noted in the literature. According to \citet{Bolmont14} and \citet{Bolmont15}, Kepler-62f and Kepler-186f may still be evolving toward a tidally locked state. \citet{Shields16} showed that, depending on whether a constant phase lag or constant time lag model is used, the rotation of Kepler-62f may or may not be synchronized within few-gigayear timescales. They also illustrate a strong tidal timescale dependence on initial spin period and orbital eccentricity. Furthermore, the ages of the systems are highly uncertain: $7\pm4$ Gyr \citep{Borucki13} or $2.34^{+2.15}_{-1.02}$ Gyr \citep{Morton16} for Kepler-62 and $4\pm0.6$ Gyr \citep{Quintana14} for Kepler-186.

The tidal synchronization timescale assuming a constant time lag tidal model follows the equation below \citep[e.g.,][]{Hut82, Heller11, Ogilvie14}:
\begin{align}
\frac{1}{\tau_{syn}} = \Big|\frac{\dot{\omega}}{\omega}\Big| = 3k_2 \Delta t |\Omega_{orb} - \omega| \frac{L_{orb}}{L_s}\frac{M_*}{M_p}\Big(\frac{R_p}{a}\Big)^5\Omega_{orb} ,
\end{align}
where $\omega$ is the planetary rotational angular velocity, $\Omega_{\rm orb}$ is the orbital angular velocity, $L_{\rm orb}$ is the orbital angular momentum, and $L_s = C \omega$ is the spin angular momentum. If the planets have the same love number and tidal time lag as the Earth, i.e. $k_2 \Delta t = 213$s \citep{Lambeck80},\footnote{The high dissipation rate of the Earth is largely due to its oceans, which may be a unique feature and unrepresentative of terrestrial planets in general.} then the characteristic tidal synchronization timescales for Kepler-62f and Kepler-186f with 24 hr rotation periods are $\sim 5.5$ and $\sim 0.3$ Gyr, respectively. 

Regardless of whether the planets are still en route to tidal locking or have completed the journey in the past, tidal interaction has influenced the history of, and may continue to affect the evolution of, the planets' obliquities. As the rotation rate of the planet evolves, the oblateness of the planet changes, which leads to a varying precession coefficient ($\alpha_p$). Thus, it is possible for the planet to move across different resonant regions during the tidal synchronization, as illustrated in Figure \ref{f:5p-mad} and \ref{f:7p-mad}. The fact that the tidal timescale for the alignment of the planetary spin axis is much longer than the obliquity variation timescales in the resonant zones means that long-term tidal effects cannot suppress short-term obliquity fluctuations. Consequently, the obliquity can still vary owing to the resonant interactions. This would be similar to the future obliquity evolution of Earth as discussed in \citet{deSurgy97}. For Kepler-62f and Kepler-186f, their respective synchronized spin periods are too long to allow resonant interactions (see Figure \ref{f:5p-mad} and \ref{f:7p-mad}). Thus, the obliquities are stable for both Kepler-62f and Kepler-186f in the synchronized stage.\footnote{A synchronously rotating body could also sustain large obliquity variations if it lies in a chaotic zone \citep{Wisdom84}.}

\bigskip

\section{Conclusions}\label{s:conclusion}

In this article, we have investigated the short-term obliquity variability of HZ planets in two multiplanet transiting systems, Kepler-62f and Kepler-186f, over a large parameter space of possible planet properties and orbital architectures allowed by observational constraints. Using {\emph N}-body simulations coupled with secular spin-orbit coupling analysis, we have shown in section \textsection \ref{s:nr} that low-obliquity regions of Kepler-62f and Kepler-186f are stable over $10$ Myr timescales while higher-obliquity regions allow small variabilities, assuming that the planets are Earth analogs (i.e. same rotation rate and interior structure, obeying Eqn.(\ref{e:alpha_earth})). 

We have also presented an analytical framework to characterize the nature of obliquity instabilities from first principles (\textsection \ref{s:ar}). The basic elements of the method are as follows:

\begin{enumerate}
\item Present the nature of obliquity instability as arising from resonant interactions between the planetary spin axis and the orbital axis \citep[e.g.,][]{Laskar93a}. Wherever there is a match between the spin axis precession and inclination oscillation frequency, obliquity variation could occur (\textsection \ref{ss:res-zones}). Given a forcing frequency, the location at which a resonance occurs can be calculated from Eqn.(\ref{e:coseps}).
\item Derive the expression for the width of the resonant zones from the modal amplitude of the forcing inclination vector (Eqn.(\ref{e:wid}), \textsection \ref{ss:res-wid}). 
\item Deduce modal frequencies and amplitudes in the orbital inclination vector of the planet of interest (e.g., Kepler-62f and Kepler-186f). This could be done in two ways:
\begin{enumerate}
    \item numerically through FT on the output of an {\emph N}-body simulation of the system (\textsection \ref{ss:num-freq}), or
    \item analytically through the Lagrange-Laplace formalism (\textsection \ref{ss:ll-th}). The only inputs required are the initial planetary system architecture (component masses and semi-major axes).
\end{enumerate}  
\end{enumerate}

When applicable, the analytical technique has the decided advantage of being a rapid, straightforward, and transparent way to compute the regions harboring resonant obliquities over a large parameter space. It provides a visualization for the behavior of obliquity variablity as these parameters vary. For Kepler-62f and Kepler-186f, we have shown good agreement between the numerical and analytical approach. 

Different planetary spin rates and orbital configurations of the Kepler systems could affect the obliquity variations. For instance, for Kepler-62f, the lower-obliquity region ($\sim 20^\circ-40^\circ$) can be unstable when the rotation period is $\sim 15-20$ hr or $\sim 60-70$ hr. For Kepler-186f, the same lower-obliquity region can be unstable when the rotation period is $\sim 240-300$ hr. For both planets, instability in the higher-obliquity regions ($\gtrsim 60^\circ$) occurs for a wider range of rotation periods ($\sim 0.3 - 3$ days). The specific values of the rotation period also depend on the properties of the assumed planetary interior structure. In general, instability in the lower-obliquity region is fine-tuned, while the higher-obliquity region can be unstable for a wider parameter space. The amplitude of variability is dependent on the mutual orbital inclination of the planets in the system. Configurations deviating from coplanarity by $\sim 3^\circ$ can already generate appreciable ($\sim 20^\circ$) fluctuations in the low-obliquity ranges.


Orbital architectures and planet properties are often difficult to measure and/or subjected to update. Our analytical approach enables us to characterize the overall obliquity variations including observational uncertainties, different planetary oblateness (which leads to different precession coefficients), extra planets, and the existence of satellites. We  find  that  the  observational uncertainties in the stellar mass and in the estimates of the planetary mass do not change our conclusion qualitatively. In investigating the impact of extra planets, we find that Jupiter and Saturn analogs can induce larger obliquity variations in the lower-obliquity range for Kepler-62f, assuming an Earth analog. The obliquity variations of Kepler-186f are not strongly affected by this specific realization of external giant planets. However, an extra planet between Kepler-186e and Kepler-186f may induce stronger obliquity variations for Kepler-186f in the low-obliquity region. Assuming rotation rates similar to that of the Earth, the existence of moons for Kepler-62f could destabilize the spin axis of Kepler-62f, but they cannot destabilize the spin axis of Kepler-186f. Long-term tidal interactions between the planet and the host star will synchronize the planetary spin axis and reduce the oblateness of the planet and its precession coefficient, moving it across resonant regions. Thus, one would expect the obliquity of the planets to vary with large amplitude during tidal synchronization, before reaching the obliquity-stable regions at synchronization.


Based on a simplified energy-balance model, \citet{Armstrong14} showed that rapid and large obliquity variability can be favorable to life by keeping a planet's global average temperature higher than it would have been otherwise, thereby systematically extending the outer edge of a host star's HZ by $\sim 20\%$. However, it is by no means clear whether large obliquity variation is necessarily beneficial to life under all circumstances. For instance, it is believed that large obliquity variation for Mars may have caused the collapse of its atmosphere and rendered Mars inhabitable \citep[e.g.,][]{Toon80, Soto12}. At the very least, obliquity variability can substantially affect transitions between multiple climate states. Recently, \citet{Kilic17} mapped out the various equilibrium climate states reached by an Earth-like planet as a function of stellar irradiance and obliquity. They find that, in this parameter space, the state boundaries (e.g. between cryo- and aqua-planets) are sharp and very sensitive to the climate history of the planet. This suggests that a variable obliquity can easily move the planet across state divisions, as well as alter the boundaries themselves, which would translate into a dramatic impact on instantaneous surface conditions and long-term climate evolution. Similarly, the dependence of surface incident flux on obliquity and eccentricity was studied by \citet{Kane17}. They found that nonzero obliquity values could potentially effect large variations in insolation flux across planetary latitudes and orbital phases. Incidentally, among their test cases was Kepler-186f. Coupling the evolution of flux maps with that of obliquity could yield further insight into climate development on exoplanets. The detailed effects on the climate due to obliquity variations still need more investigation. While atmospheric modeling is beyond the scope of this study, our work can help provide input parameters to existing global climate models (GCMs) as another factor influencing the habilitability in a multiplanet system.

\bigskip
\section*{Acknowledgments}
The authors would like to thank Konstantin Batygin, Guillermo Torres, Jack Lissauer and Billy Quarles for helpful discussions. We are also grateful to Jennifer Yee, Matthew Holman, and the anonymous referee for providing detailed feedback on our draft. This work and G.L. were supported in part by the Harvard William F. Milton Award. Y.S. is supported by a Doctoral Postgraduate Scholarship from the Natural Science and Engineering Research Council (NSERC) of Canada.

\end{document}

%% file: System_pars.tex
\setlength{\tabcolsep}{3pt}
\begin{table}[htb]
\begin{center}
\caption{Planetary System Parameters Used in This Work}
\label{t:sys-pars}
\begin{tabular}{lrcccc}
\hline\hline
Planet & Period, & Inclination, & Radius, & Mass, \\
 & $P_{\rm orb}$ (days) & $i_{\rm LoS}$ & $R_p$ ($R_\earth$) & $M_p$ ($M_\earth$) \\
\hline
\multicolumn{5}{c}{\bf{Kepler-62}}\\
\hline
b & 5.715 & 89.2 $\pm$ 0.4 & 1.31 $\pm$ 0.04 & $2.2^{+1.6}_{-0.7}$ \\
c & 12.442 & 89.7 $\pm$ 0.2 & 0.54 $\pm$ 0.03 & $0.11^{+0.05}_{-0.04}$ \\
d & 18.164 & 89.7 $\pm$ 0.3 & 1.95 $\pm$ 0.07 & $4.8^{+3.8}_{-1.9}$ \\
e & 122.387 & 89.98 $\pm$ 0.02 & 1.61 $\pm$ 0.05 & $3.6^{+2.4}_{-1.3}$ \\
f & 267.291 & 89.90 $\pm$ 0.03 & 1.41 $\pm$ 0.07 & $3.3^{+2.3}_{-0.6}$ \\

\hline
\multicolumn{5}{c}{\bf{Kepler-186}}\\
\hline
b & 3.887 & 88.9 $\pm$ 0.7 & 1.19 $\pm$ 0.08 & $1.7^{+1.2}_{-0.6}$ \\
c & 7.267 & 89.3 $\pm$ 0.4 & 1.38 $\pm$ 0.09 & $2.5^{+2.1}_{-0.8}$ \\
d & 13.343 & 89.4 $\pm$ 0.3 & 1.55 $\pm$ 0.11 & $3.4^{+2.4}_{-1.2}$ \\
e & 22.408 & 89.7 $\pm$ 0.2 & 1.41 $\pm$ 0.10 & $2.7^{+1.9}_{-1.0}$ \\
f & 129.946 & $89.96^{+0.04}_{-0.10}$ & 1.17 $\pm$ 0.11 & $1.7^{+1.1}_{-0.3}$ \\

\hline\hline
\end{tabular}
\end{center}
Note. $P_{\rm orb}$, $i_{\rm LoS}$ and $R_p$ for Kepler-62 are obtained from \citet{Borucki13}. $P_{\rm orb}$ for Kepler-186 are obtained from \citet{Quintana14}. $i_{\rm LoS}$ and $R_p$ for Kepler-186 are calculated based on the updated stellar mass and radius from \citet{Torres15} and the planetary system parameters (impact parameters, orbital periods, and transit depths) measured by \citet{Quintana14}. The planetary masses for both Kepler-62 and Kepler-186 are estimated using the publicly available code {\tt{Forecaster}} of \citet{ChenKip17}.

\end{table}
\setlength{\tabcolsep}{6pt}